\documentclass[11pt,a4paper]{article}
\pdfoutput=1

\usepackage{amsmath,amssymb,bm,float,graphics,mathrsfs,color,jcappub}

\newcommand{\bc}{\begin{center}}
\newcommand{\ec}{\end{center}}
\newcommand{\beq}{\begin{equation}}
\newcommand{\eeq}{\end{equation}}
\newcommand{\Tr}{\mathrm{Tr}}
\newcommand{\PD}[2]{\dfrac{\partial #1}{\partial #2}}
\newcommand{\al}[1]{\begin{align} #1 \end{align}}
\newcommand{\ave}[1]{\left\langle #1 \right\rangle}
\newcommand{\bi}{\begin{itemize}}
\newcommand{\ei}{\end{itemize}}
\def\mcO{{\mathcal O}}
\def\hatn{\hat{\bm{n}}}
\def\vs#1{\vspace{#1\baselineskip}}
\def\B#1{\textbf{#1}}
\def\rom#1{_{\mathrm{#1}}}
\def\l{\ell}
\def\bl{\bm\l}
\def\Ob{\Omega_{\rm b}}
\def\Om{\Omega_{\rm m}}
\def\OL{\Omega_{\Lambda}}
\def\mnu{\sum m_{\nu}}

\title{Probing dark energy and neutrino mass from upcoming lensing experiments of CMB and galaxies}

\author[a]{Toshiya Namikawa}
\author[a,b]{Shun Saito}
\author[c,d]{and Atsushi Taruya}

\affiliation[a]{Department of Physics, Graduate School of Science, The University of Tokyo, Tokyo 113-0033, Japan}
\affiliation[b]{Department of Astronomy, 601 Campbell Hall, University of California Berkeley, CA 94720, USA}
\affiliation[c]{Research Center for the Early Universe, School of Science, The University of Tokyo, Bunkyo-ku, Tokyo 113-0033, Japan}
\affiliation[d]{Institute for the Physics and Mathematics of the Universe, The University of Tokyo, Kashiwa, Chiba 277-8568, Japan}

\emailAdd{namikawa@utap.phys.s.u-tokyo.ac.jp}

\abstract{
We discuss the synergy of the cosmic shear and CMB lensing experiments to simultaneously constrain the neutrino mass and dark energy properties. Taking fully account of the CMB lensing, cosmic shear, CMB anisotropies, and their cross correlation signals, we clarify a role of each signal, and investigate the extent to which the upcoming observations by a high-angular resolution experiment of CMB and deep galaxy imaging survey can tightly constrain the neutrino mass and dark energy equation-of-state parameters. Including the primary CMB information as a prior cosmological information, the Fisher analysis reveals that the time varying equation-of-state parameters, given by $w(a)=w_0+w_a(1-a)$, can be tightly constrained with the accuracies of $5\%$ for $w_0$ and $15\%$ for $w_a$, which are comparable to or even better than those of the stage-III type surveys neglecting the effect of massive neutrinos. In other words, including the neutrino mass in the parameter estimation would not drastically alter the Figure-of-Merit estimates of dark energy parameters from the weak lensing measurements. For the neutrino mass, a clear signal for total neutrino mass with $\sim0.1$ eV can be detected with $\sim2\sigma$ significance. The robustness and sensitivity of these results are checked in detail by allowing the setup of cosmic shear experiment to vary as a function of observation time or exposure time, showing that the improvement of the constraints very weakly depends on the survey parameters, and the results mentioned above are nearly optimal for the dark energy parameters and the neutrino mass. 
}

\begin{document}

\maketitle

\section{Introduction} \label{sec1}

Cosmological observations in the last decade have successfully led to the establishment of standard cosmological model, and the energy composition in the Universe is well-determined with an accuracy less than $\sim10\%$ level. In particular, the observations of cosmic microwave background (CMB) and large-scale structure confirmed that our universe is described by a flat $\Lambda$ cold dark matter ($\Lambda$CDM) model, in which the Universe is filled with $\sim30$ \% of cold dark matter and baryon, and $\sim70$ \% of the dark energy \cite{Komatsu:2010fb,Amanullah:2010vv}. Together with the first discovery of the late-time cosmic acceleration by the observations of Type Ia supernovae \cite{Riess98,Perl99}, we are sure that the Universe at the present day is dominated by the unknown energy components. 

Taking seriously account of our current understanding and lack of our knowledge of the Universe, the cosmology in the coming decade should focus on more profound or even more advanced issues; origin and nature of the dark energy and dark matter, and in other words, a more accurate description of our Universe beyond a flat $\Lambda$CDM model. The answers to these questions are important, and would shed light on new physics beyond our current knowledge. 

In this paper, we consider the prospects of upcoming observations for tightly constraining the nature of the dark energy and the neutrino mass. The theoretical understanding of the nature of the dark energy is still limited, and the cosmological observations are the only way to reveal the dynamical properties of the dark energy. For this purpose, in the literature, the dark energy are often parametrized with its equation-of-state, i.e., $p=w\rho$, allowing a time dependence of $w$. Combining several complementary observations, one tries to determine and/or constrain $w$ as a function of time. On the other hand, determination of the neutrino mass is one of the important subject in elementary particle physics, and is the key to understand the physics beyond the standard model of particle physics. In cosmology, the massive neutrinos play a role of hot dark matter, and they can alter the cosmic expansion history as well as the growth of large-scale structure \cite{BES80,Lesgourgues:2006nd}. An interesting point here is that the cosmological observations can give a tight bound on the sum of neutrino masses, and have currently achieved the constraint $\mnu<0.2-1.0 $eV (95\% C.L.) (e.g., Refs.~\cite{Reid:2009nq,Saito:2010pw,I09}), which is comparable or even stronger than the constraint obtained from the ground-based experiment using $\beta$-decay \cite{MSTV04}. Thus, a pursuit of constraining or measuring neutrino mass from upcoming observations is scientifically fruitful and at least complementary to ground-based experiments. 

Here, we are specifically concerned with the weak lensing effects on the CMB and galaxies, as representative cosmological probes for dark energy properties and neutrino mass. The main advantage of the weak lensing effect is to directly map the large-scale structure, and we can simultaneously extract the information on the cosmic expansion and structure growth imprinted on the lensed CMB map and galaxy images. In particular, the weak lensing of galaxies, referred to as the cosmic shear, is expected to be powerful to constrain the dark energy and total neutrino mass from future observations, and several works on the forecast study have been done \cite{SK04,HTW06,Alb06,Abazajian:2002ck,Kitching:2008dp,DeBernardis:2009di}. On the other hand, weak lensing of CMB, often quoted as CMB lensing, is another sensitive probe of the matter density fluctuations at higher redshift, and a signature of lensed CMB has been recently detected with a high significance through cross correlation with galaxy clustering \cite{Smith:2007rg,Hirata08}, and directly through the CMB power spectrum \cite{Calabrese:2008rt,Das:2010ga}. Future CMB observations such as Planck will definitely detect CMB lensing, and a synergy with other cosmological probes would help to give a tight constraint on dark energy and neutrino mass. 

As a representative example, we discuss a prospect and a synergy between specific upcoming CMB lensing and cosmic shear experiments; cosmic shear survey with the Hyper Suprime-Cam (HSC) mounted on Subaru telescope \cite{HSC}, and the CMB lensing experiment with the Planck and Atacama Cosmology Telescope with new polarization sensitive receiver (ACTPol) \cite{Niemack:2010wz}. The cosmic shear survey with HSC plans to observe galaxy images over the sky of $2,000$ square degree, with $35$ galaxies per square arcminute of the mean redshift $z_m\sim1$. On the other hand, ACTPol has enough sensitivity to measure the CMB lensing even at very small scales with higher signal-to-noise ratio than Planck. An important characteristic in these two experiments is that survey regions of these observations will be overlapped each other, capable of dealing with cross correlation study. 

Motivated by these situations, in this paper, we address the feasibility and complementarity of the upcoming lensing experiments. Based on the Fisher analysis, we quantitatively investigate how well we can constrain the dark energy and total neutrino mass, taking fully account of the CMB lensing, cosmic shear, CMB anisotropies, and their cross correlations. To our knowledge, this is the first work to consider full cross correlation signals between the two lensing signals, including the effects of both the dark energy and the neutrino mass. There have been several forecast studies on both the CMB lensing and cosmic shear, but these are restricted to specific models such as the early dark energy \cite{HSCS09} or those neglecting the massive neutrinos \cite{Hu01}. We here discuss a role of each signal and the influence of parameter degeneracies in constraining the neutrino mass and dark energy properties. Further, dependence of the forecast results on the survey setup is investigated in detail. In this respect, the analysis in this paper includes a general result not only valid for the specific lensing surveys with HSC and ACTPol, but also applicable to other lensing experiments. 

This paper is organized as follows. In Sec.~\ref{sec2}, we briefly review the CMB lensing and cosmic shear. In Sec.~\ref{sec3}, we summarize the Fisher matrix formalism, and describe the canonical setup of the CMB lensing and cosmic shear experiments with Planck, ACTPol and HSC. Then, in Sec.~\ref{sec4}, the signal-to-noise ratio and forecast results for parameter estimations are presented in detail, focusing on the dark energy and total neutrino mass. In Sec.~\ref{sec5}, to see the sensitivity of the forecast results to the details of the survey setup, we allow to vary the setup of cosmic shear experiment, and examine how the constraints on neutrino mass and dark energy properties are changed. Finally, Sec.~\ref{sec6} is devoted to summary and conclusion.

Throughout the paper, we calculate the power spectra for a fiducial set of cosmological parameters in Table \ref{Fid}, i.e., the density parameter of baryon $\Ob h^2$, of matter $\Om h^2$, dark energy density $\OL$, scalar spectral index $n\rom{s}$, and scalar amplitude $A\rom{s}$ at $k=0.002$ Mpc$^{-1}$, reionization optical depth $\tau$, total neutrino mass $\mnu$, and dark energy equation-of-state parameters, $w_0$ and $w_a$, with the functional form of $w(a)=w_0+w_a(1-a)$ (e.g., \cite{Chevallier:2000qy}). 

\begin{table}
\bc
\begin{tabular}{ccccccccc} \hline 
$\Ob h^2$ & $\Om h^2$ & $\OL$ & $n_{\rm s}$ & $A_{\rm s}\times 10^9$ & $\tau$ & $\mnu$ & $w_0$ & $w_a$ \\ \hline 
$0.022$ & $0.13$ & $0.72$ & $0.96$ & $2.4$ & $0.086$ & $0.1$ eV & $-1.0$ & $0.0$ \\ \hline
\end{tabular}
\ec
\vs{-1}
\caption{
Fiducial values of cosmological parameters used in this paper. These values are favored in the WMAP results \cite{Komatsu:2010fb}. We assume the flat $\Lambda$CDM model with three massive neutrinos.
}
\label{Fid}
\end{table}

\section{Weak gravitational lensing of CMB and galaxies} \label{sec2}

\subsection{Power spectra of weak gravitational lensing}

Primary CMB temperature, $\Theta(\hatn)$, and polarization fields, $E(\hatn)$ and $B(\hatn)$, are distorted due to the large scale structure between the last scattering surface and us (see \cite{LC06} for a review). The weak lensing effect on the CMB fields is characterized by the deflection vector, $\bm{d}$ such as 
\beq 
	\tilde{\Theta}(\hatn) = \Theta(\hatn+\bm{d}) \,, 
\eeq 
where $\tilde{\Theta}$ denotes the lensed temperature field, and the same is as well with the case of polarization fields. Note that lensed $B$-mode polarization can be generated by lensed distortion of primary $E$-mode \cite{Zaldarriaga:1998ar}. The deflection angle due to the gravitational lensing is obtained by solving the geodesic equation, yielding 
\beq
	\bm{d}(\hatn) = -2\int_0^{\chi_*}d\chi\,\frac{\chi_*-\chi}{\chi_*\chi} \, 
		\nabla_{\hatn}\psi(\chi\hatn,\eta_0-\chi), \label{def} 
\eeq
where $\chi_*$ denotes the comoving radial distance to the last scattering surface, $\eta_0$ is the conformal time at present, and $\psi(\bm{r},\eta)$ is the Newton potential. Note that we set units for the light velocity. The deflection field thus traces the gravitational field produced by the structure between the last scattering surface and us. Using Eq.~\eqref{def}, a statistical quantity of our interest, i.e., the angular power spectrum of the deflection angle, $C_\l^{dd}$, is theoretically computed as \cite{Hu01}
\beq
	C_\l^{dd}=\frac{2}{\pi}\int \frac{dk}{k^2}\, P\rom{init}(k) [\Delta^d_\l(k)]^2 
	\,, \label{dd}
\eeq
where $P\rom{init}(k)$ is the matter power spectrum at an early time. The function $\Delta^d_\l(k)$ is given by 
\beq 
	\Delta^d_\l(k) = \sqrt{\frac{(\l+1)!}{(\l-1)!}}\, 3\Om H_0^2
		\int_0^{\chi_*}d\chi\,\frac{\chi_*-\chi}{\chi_*\,\chi}\,\frac{D(k; z(\chi))}{a(\chi)}\,j_\l(k\chi)
	\,, \label{kernel-d} 
\eeq 
with $\Om$ and $H_0$ being the density parameter of mass and the Hubble parameter at the present time, respectively. The function $D(k; z)$ represents the growth factor, defined by the square root of the ratio of matter power spectra, $D(k; z)\equiv\sqrt{P(k;z)/P\rom{init}(k)}$. Note that the scale-dependent growth naturally arises from the non-linear gravitational evolution and free-streaming suppression by the massive neutrinos (see \ref{sec2.2}). 

The deflection field can be efficiently reconstructed from the non-Gaussian nature in the lensed CMB statistics \cite{HO02,OH03}. In the flat-sky approximation, the estimator for the deflection field is given by \cite{HO02} 
\beq 
	\hat{d}^{XY}(\bl)\equiv \frac{A^{XY}(\bl)}{\l}
		\int\frac{d^2\bl'}{(2\pi)^2}\tilde{X}(\bl')\tilde{Y}(\bl-\bl')g_{XY}(\bl,\bl')
	\,, 
\eeq
where the subscripts $\tilde{X}$ and $\tilde{Y}$ denote the lensed temperature and polarization fields, i.e., $\tilde{\Theta}$, $\tilde{E}$, $\tilde{B}$, and $A^{XY}(\bl)$ and $g_{XY}(\bl,\bl')$ denote the normalization and the optimal weight function, respectively (see \cite{HO02} for explicit expressions for $A^{XY}(\bl)$ and $g_{XY}(\bl,\bl')$). The essence of this quadratic estimator is that information from different scales is used to reconstruct the deflection field. The optimal weight is chosen so as to minimize the lensing reconstruction error, $N_\l^{dd}$. If one considers the temperature field only, for an illustrative example, the error $N_\l^{dd}$ is estimated as \cite{HO02}
\beq 
	N^{dd}(\bl) = \l^2\left[\int\frac{d^2\bl'}{(2\pi)^2}\frac{[(\bl-\bl')\cdot\bl'
		C_{|\bl-\bl'|}^{\Theta\Theta} + \bl\cdot\bl' C_\l^{\Theta\Theta}]^2}
		{2(C_\l^{\Theta\Theta}+N_\l^{\Theta\Theta})(C_{|\bl-\bl'|}^{\Theta\Theta}
		+N_{|\bl-\bl'|}^{\Theta\Theta})}\right]^{-1}
	\,, 
\eeq
where the quantity $N_\l^{\Theta\Theta}$ is the temperature noise power spectrum (see Sec.~\ref{sec3}). In this paper, we compute $N_\l^{dd}$ following the technique developed in \cite{OH03} in which the temperature and polarization fields are optimally combined in the full-sky treatment. 

On the other hand, in a weak lensing survey of galaxies, we first measure the ellipticity of each galaxy image and then estimate the shear field $\gamma(\hatn)$ \cite{Munshi:2006fn}. Although the shear field is projected onto a two-dimensional sky, the inclusion of redshift information significantly improves the sensitivity to the cosmological distance and structure growth, which leads to tighter constraints on the dark energy. This technique is referred to as the tomography \cite{Hu99}, and with the photometric redshift samples divided into several subsamples binned with redshifts, we obtain various combinations of the cross shear spectrum, e.g., $C_\l^{\gamma_i\gamma_j}$ for cross-spectrum between $i$-th and $j$-th redshift bins. Given redshift distribution of source galaxies in $i$-th, $n_i(z)$, the shear power spectrum is calculated from \cite{Hu01} 
\beq
	C_\l^{\gamma_i\gamma_j}=\frac{2}{\pi}\int\frac{dk}{k^2}\,P\rom{init}(k)\Delta_\l^{\gamma_i}(k)\Delta_\l^{\gamma_j}(k) 
	\,, \label{SS}
\eeq
where $\Delta_\l^{\gamma_i}(k)$ is given by 
\beq
	\Delta_\l^{\gamma_i}(k) = \sqrt{\frac{(\l+2)!}{(\l-2)!}}\,\frac{3\Om H_0^2}{2}\,
		\int_0^{\infty}dz_s\,\frac{n_i(z_s)}{\bar{n}_i}\int_0^{\chi(z_s)}d\chi\,
		\frac{\chi(z_s)-\chi}{\chi(z_s)\,\chi}\,\frac{D(k;z(\chi))}{a(\chi)}\, j_\l(k\chi) 
	\,. \label{kernel-S} 
\eeq
The quantity $\bar{n}_i$ is the average number density per square arcminute in $i$-th bin, defined as 
\beq 
	\bar{n}_i = \int_0^{\infty}n_i(z)dz \,.  
\eeq
Note that apart from the prefactor $(\l+2)(\l-1)/4$, the expression \eqref{SS} with \eqref{kernel-S} is reduced to that of the CMB lensing \eqref{dd} if we choose $n(z)=\delta_D(z-z_*)$, where $\delta_D$ represents the Dirac's delta function and $z_*$ means the redshift at the last scattering surface. Thus, the shear power spectrum $C_\l^{\gamma\gamma}$ generically has a larger power at small scales ($\l\gg1$) than the CMB lensing spectrum $C_\l^{dd}$. 

In addition, for a survey region overlapping between the cosmic shear and CMB lensing experiments, the cross-correlation signal between the deflection and shear fields can be obtained. Since both the CMB lensing and cosmic shear signals originated from the large-scale structure, the cosmological information from the two lensing measurements is statistically correlated. Therefore, the cross-correlation signal must be properly taken into account in cosmological analysis. The cross-power spectrum between the CMB deflection and galaxy shear field is predicted as \cite{Hu01}
\beq 
	C_\l^{d\gamma_i} = \frac{2}{\pi}\int \frac{dk}{k^2}\, P\rom{init}(k)\,\Delta^d_\l(k)\Delta_\l^{\gamma_i}(k)
	\,. \label{dS} 
\eeq

Finally, another important cosmological signal is obtained through the late-time integrated Sachs-Wolfe (ISW) effect, which results from the late-time decay of the gravitational potential fluctuations induced by the departure of cosmic expansion from that of the Einstein-de-Sitter universe. The ISW effect produces a non-vanishing cross-correlation between the CMB temperature and weak lensing fields, and its characteristic signal appears at large angular scales. The angular power spectrum of CMB temperature and weak lensing field $X$ (i.e., $X=d$ or $\gamma_i$) becomes \cite{Hu01}
\beq
	C_\l^{\Theta X} = \frac{2}{\pi}\int \frac{dk}{k^2}\,P\rom{init}(k)\Delta_\l^{\rm ISW}(k)\Delta_\l^X(k)
	\,, 
\eeq
where the kernel $\Delta_\l^{\rm ISW}(k)$ is given by
\beq
	\Delta_\l^{\rm ISW}(k) = 3\Om H_0^2\int_0^{z_*} dz \frac{d}{dz}\left\{\frac{D(k;z)}{a}\right\}\, j_\l(k\chi(z))
	\,. 
\eeq

All the spectra presented above include information on the evolution of the matter power spectrum in a somewhat complicated manner. In next subsection, we will discuss the cosmological parameter dependence of these power spectra, especially focusing on neutrino mass and dark energy properties. Throughout this paper, all the power spectra are computed with the modified version of CMB Boltzmann code, CAMB \cite{CAMB}, with the fiducial cosmological parameters given in Table \ref{Fid}. For nonlinear matter power spectrum relevant to the auto and cross power spectra of cosmic shear and deflection fields, we adopt the fitting formula proposed by Smith et al. (2003) \cite{Smith03}. 

\subsection{Cosmological information from CMB lensing and cosmic shear} \label{sec2.2}

\begin{figure}
\bc
\includegraphics[width=16cm,clip]{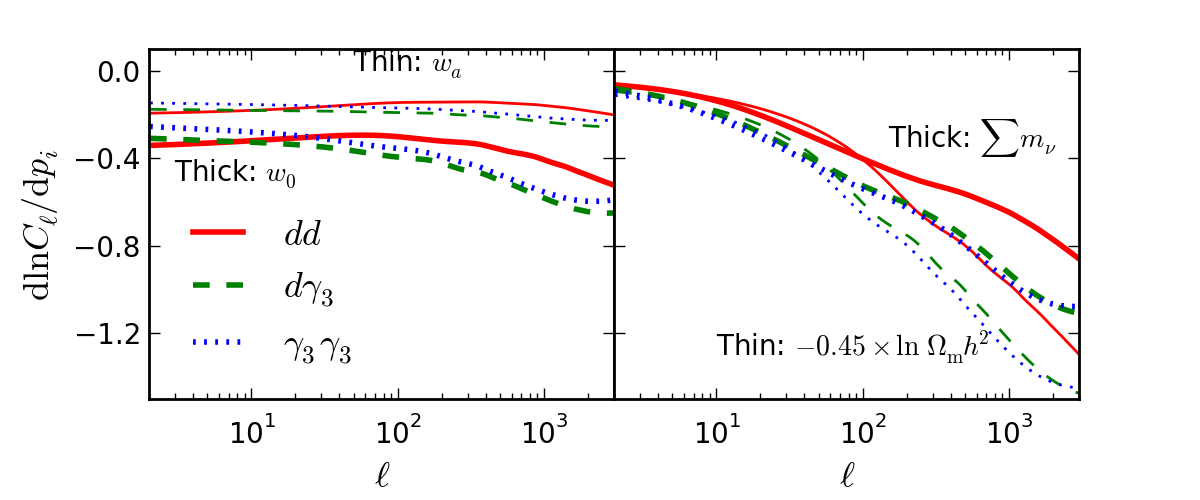} \vs{-2}
\ec 
\caption{
Logarithmic derivatives of the lensing power spectra, $C_\l^{dd}$, $C_\l^{\gamma_3\gamma_3}$ and $C_\l^{d\gamma_3}$, with respect to $w_0$ and $w_a$ (left), and $\mnu$ and $\Om h^2$ (right). The derivatives are evaluated around the fiducial set of cosmological parameters summarized in Table \ref{Fid}. To compute the derivative of cosmic shear spectra, the galaxies are divided into the three redshift subsamples as described in Sec.~\ref{sec3}, and we especially plot the results for the third subsample (i.e., $1.5<z$). Note that to see similarity of the parameter dependence, the result of $d\ln C_\l/d\ln\Om h^2$ is multiplied by $-0.45$.
}
\label{deriv}
\end{figure}

To see how the lensing spectra depend on the cosmological parameters, in Fig.~\ref{deriv}, we plot the logarithmic derivatives of angular spectra  $C_\l^{dd}$, $C_\l^{\gamma_3\gamma_3}$ and $C_\l^{d\gamma_3}$ with respect to the dark energy equation-of-state parameters $w_0$, $w_a$ (left), and the neutrino mass $\mnu$ (right). Here, the quantity $\gamma_3$ represents the shear field estimated from the galaxy subsample in the third redshift bin, whose precise meaning will be specified later (see Sec.~\ref{sec3}). Note that the cosmological dependence of the other shear fields from different redshift bins is qualitatively the same. For comparison, we also show the dependence on matter density, i.e., the logarithmic derivative $d\ln C_\l/d\ln\Om h^2$.  

The influence of dark energy on the lensing power spectra basically comes from the growth factor of matter fluctuations, $D(k;z)$, and because of the homogeneous nature of the dark energy, its effect on growth factor is nearly independent of scales. Thus, the resultant variation of angular spectra with respect to the dark energy equation-of-state parameters is almost scale-independent. On the other hand, massive neutrinos lead to a characteristic suppression on the growth of large scale structure below the free-streaming scale. Since the characteristic wavenumber of the free-streaming scale becomes large as increasing the neutrino mass, as a consequence, the logarithmic derivative of lensing spectra can be negative, and decreases as increasing multipoles. That is, the amplitude of lensing spectra is reduced with the effect of neutrino mass at small scales, and with a high angular resolution lensing experiment, we can detect a clear signature of free-streaming suppression. 

These scale-dependent or scale-independent natures of the lensing power spectra are, in principle, very powerful to constrain the neutrino mass and dark energy. However, we note here that there exist some parameters that exhibit a similar scale-(in)dependence, which can be the source of parameter degeneracy. As shown in Fig.~\ref{deriv}, the logarithmic derivative of the lensing spectra with respect to the quantity $\ln\Om h^2$ gives a similar trend to that of the neutrino mass parameter, $d\ln C_\l/d\mnu$, and thus the density parameter $\Om h^2$ can mimic a scale-dependent suppression by massive neutrino. Also, the equation-of-state parameters $w_0$ and $w_a$ can put a similar scale-independent behavior in the lensing spectra, and it seems difficult to discriminate between each other even if we use the lensing information at small scales. Note that the primary CMB information alone is insensitive to the dark energy properties and it exhibits some degeneracies, e.g., between $\ln\Om h^2$ and $\OL$. Hence, even combining the CMB data, there may remain a sizable amount of degeneracies among $w_0$ and $w_a$, $\mnu$ and $\Om h^2$. In this sense, the degree of improvement for parameter constraints seems rather non-trivial. For more quantitative aspect of the parameter estimation, we will proceed to the Fisher analysis. 

\section{Fisher matrix formalism} \label{sec3}

In this section, we summarize the Fisher matrix formalism used in the subsequent analysis, and describe the canonical setup for CMB and cosmic shear experiments, for which forecast results are presented in Sec.~\ref{sec4}. 

Given the angular power spectra theoretically parametrized by a set of parameters $\vec{p}$, the cosmological information on these parameters obtained from the combination of several experiments is quantified by the Fisher matrix \cite{TTH97}:
\beq 
	F_{ij} = \sum_{\l=2}^{\l\rom{max}}\frac{2\l+1}{2}f\rom{sky}
		\Tr\left(\bm{C}^{-1}_\l(\vec{p})\PD{\bm{C}_\l}{p_i}(\vec{p})
		\bm{C}^{-1}_\l(\vec{p})\PD{\bm{C}_\l}{p_j}(\vec{p})\right)
	\,, \label{FisMat} 
\eeq
where the quantity $\bm{C}_\l$ represents the covariance matrix for angular power spectra given below, $p_i$ is a cosmological parameter which we want to estimate, and the $f\rom{sky}$ is the sky coverage. Using the Fisher matrix, the $1\sigma$ (68\%C.L.) statistical uncertainties for the cosmological parameter $p_i$ marginalized over other parameters, $\sigma(p_i)$, is estimated as $\sigma(p_i)=\sqrt{\{\bm{F}^{-1}\}_{ii}}$. Also, the statistical correlation or degeneracy between parameters $p_i$ and $p_j$ can be deduced from the off-diagonal component of inverse Fisher matrix $\{\bm{F}^{-1}\}_{ij}$, and is quantified by defining the correlation coefficient, $r(p_i,p_j)=\{\bm{F}^{-1}\}_{ij}/\{\{\bm{F}^{-1}\}_{ii}\{\bm{F}^{-1}\}_{jj}\}^{1/2}$. It is worth noting that Eq.(\ref{FisMat}) relies on the assumption of the Gaussianity for likelihood function, which is incorrect in some situations. In particular, cosmic shear is sensitive to the nonlinear gravitational evolution, which leads to the non-Gaussian covariance. However, the actual impact of non-Gaussian covariance on the parameter estimation is turned out to be insignificant, and it degrades at most few percent level \cite{Takada:2008fn,Takahashi:2009ty}. Although this does not imply the validity of the error estimation with Gaussian likelihood function (for example, in the case of neutrino mass, the condition, $\mnu>0$, leads to non-Gaussian error \cite{Perotto:2006rj}), we adopt Eq.~\eqref{FisMat} to investigate the potential power of weak lensing experiments.  

Throughout the paper, we specifically consider Planck, ACTPol and HSC as the representative CMB and cosmic shear experiments, from which we can obtain the temperature ($\Theta$) and (E-mode) polarization ($E$) data for the primary CMB anisotropies, the deflection angle ($d$) data for the CMB lensing, and the shear field data of galaxies divided into several redshift bins ($\gamma_i$; $i=1,\cdots,N$). Then, the full covariance matrix, $\bm{C}_\l$, is written in the form as 
\beq 
	\bm{C}_\l = \begin{pmatrix} 
		C^{\Theta\Theta}_\l+N^{\Theta\Theta}_\l & C^{\Theta E}_\l & 
		C^{\Theta d}_\l & C^{\Theta\gamma_1}_\l & \cdots & C^{\Theta\gamma_n}_\l \\
		C^{\Theta E}_\l & C^{EE}_\l+N^{EE}_\l & 0 & 0 & \cdots & 0 \\
		C^{\Theta d}_\l & 0 & C^{dd}_\l+N^{dd}_\l & C^{d\gamma_1}_\l & \cdots & C^{d\gamma_n}_\l \\
		C^{\Theta\gamma_1}_\l & 0 & C^{d\gamma_1}_\l & 
		C^{\gamma_1\gamma_1}_\l+N^{\gamma_1\gamma_1}_\l & \cdots & C^{\gamma_1\gamma_n}_\l 
		\\ \vdots & 0 & \vdots & \vdots & \ddots & \vdots \\
		C^{\Theta\gamma_n}_\l & 0 & C^{d\gamma_n}_\l & C^{\gamma_n\gamma_1}_\l & \cdots & 
		C^{\gamma_n\gamma_n}_\l+N^{\gamma_n\gamma_n}_\l \\ \end{pmatrix}
	\,. \label{Cov-all} 
\eeq
Here, $N_\l^{XY}$ is the noise power spectrum. The amplitude and shape of the noise spectra $N_\l^{XY}$ depends on the survey design for the CMB and lensing experiments, which will be discussed below. 

For the CMB experiment, Planck, we use seven frequency channels for temperature and polarization observations, assuming the sky coverage of $f_{\rm sky}=0.65$. As for the ground-based experiment, ACTPol, single frequency channel with $\nu=148$GHz is used for high-resolution observation, and we assume $f_{\rm sky}=0.1$ to enhance the CMB lensing information \footnote{This setup is somewhat different from the original proposals of Ref.~\cite{Niemack:2010wz}, but we here keep it to investigate the potential power of high resolution experiment.}. When estimating the temperature and polarization power spectra from ACTPol, we combine the Planck data of $\l<700$, in order to remedy a large uncertainty at large angular scales arising from the atmospheric temperature fluctuations. 

In both Planck and ACTPol, the dominant noise source for temperature and polarization observations may be the photon shot noise. The noise power spectra are then expressed as
\al{
	N_\l^{XX} &= \left[\sum_\nu(N_{\l,\nu}^{XX})^{-1}\right]^{-1}; & N_{\l,\nu}^{XX} 
		&\equiv\left(\frac{\sigma_\nu\theta_\nu}{T\rom{CMB}}\right)^2\exp\left[\frac{\l(\l+1)\theta_\nu^2}{8\ln 2}\right] 
	\,, \label{noise}
}
with $T\rom{CMB}=2.7$K being mean temperature of CMB. Here, the quantity $\theta_\nu$ is the beam size, and $\sigma_\nu$ represents the sensitivity of each channel to the temperature $\sigma_{\nu,T}$ or polarization $\sigma_{\nu,P}$, depending on the power spectrum of temperature ($X=\Theta$) or polarization ($X=E$ or $B$). Specific values for these quantities are summarized in Table \ref{Instru}. Note that the foregrounds contamination in the CMB data would be additional noise source for lensing experiments, and it could not only degrade but also bias the cosmological constraints. Recent estimate by Ref.~\cite{Verde:2005ff} suggests that the foregrounds contribution to the angular power spectrum would become less than the instrumental noise, if foregrounds can be successfully subtracted at a 1\% level. Although this is still challenging and a more elaborative study is necessary for the foreground removal, we here ignore the effect of foreground contamination in order to explore the potential and complementarity of the lensing experiment.

For the CMB lensing, photon shot noise given by Eq.~(\ref{noise}) is also the dominant noise contribution, but this time, we must further consider the errors associated with reconstruction technique. We adopt the optimal quadratic estimator proposed by Ref.~\cite{OH03}, and the temperature, $E$- and $B$-mode polarization are used to estimate the deflection angle. The noise power spectrum $N_\l^{dd}$ is computed based on the expression (42) in Ref.~\cite{OH03}. 

\begin{table}
\bc
\label{Instru}
\begin{tabular}{cccccc} \hline 
Experiment & $f\rom{sky}$ & $\nu$ [GHz] & $\theta_\nu$ [arcmin] & $\sigma_{\nu,T}$ [$\mu$K/pixel] & $\sigma_{\nu,P}$ [$\mu$K/pixel] 
\\ \hline 
Planck \cite{Planck} & 0.65 & 30 & 33 & 4.4 & 6.2 \\ 
 & & 44 & 23 & 6.5 & 9.2  \\ 
 & & 70 & 14 & 9.8 & 13.9 \\
 & & 100 & 9.5 & 6.8 & 10.9 \\
 & & 143 & 7.1 & 6.0 & 11.4 \\
 & & 217 & 5.0 & 13.1 & 26.7 \\
 & & 353 & 5.0 & 40.1 & 81.2 \\ \hline 
ACTPol \cite{ACTPol} & 0.1 & 148 & 1.4 & 3.6 & 5.0 \\ \hline 
\end{tabular}
\ec
\vs{-1}
\caption{
Assumed experimental specifications for the Planck and ACTPol. The quantity $\theta_\nu$ is the beam size, and $\sigma_\nu$ represents the sensitivity of each channel to the temperature $\sigma_{\nu,T}$ or polarization $\sigma_{\nu,P}$, depending on the power spectrum of temperature ($X=\Theta$) or polarization ($X=E$ or $B$). The quantity $\nu$ means a channel frequency. 
}
\end{table}

On the other hand, canonical setup for cosmic shear survey with HSC discussed here roughly match the survey plan proposed in Ref.~\cite{HSC}. We consider the deep imaging survey with area $2,000$ deg$^2$ and mean redshift $z_m=1$, assuming the redshift distribution of galaxies given by 
\beq 
	n(z) = \frac{3 N_g}{2 z_0^3}\,z^2\,\exp\left[-\left(\frac{z}{z_0}\right)^{1.5}\right]
	\,. \label{eq:ngal_z}
\eeq
with the mean number density of galaxies, $N_g=35$ arcmin$^{-2}$. Note that $z_0$ is related to the mean redshift through $z_0=0.69 z_m$. We divide the whole galaxy sample into the three redshift subsamples (i.e., $N=3$) for lensing tomography, and use the auto and cross power spectra between different redshift bins; $0<z<0.7$, $0.7<z<1.5$, and $1.5<z$. Note that for lensing tomography based on the photometric redshift technique, the uncertainty arising from the photometric redshift error is crucial for the cosmological analysis. To mimic this effect, we suppose that the photometric redshift estimates are distributed as a Gaussian with RMS fluctuation $\sigma(z)$. Then the actual redshift distribution for $i$-th galaxy subsample becomes \cite{HS04}
\beq 
	n_i(z) = \frac{1}{2}\,n(z)\left[{\rm erfc}\left(\frac{z_i-z}{\sqrt{2}\sigma(z)}\right)-{\rm erfc}
		\left(\frac{z_{i+1}-z}{\sqrt{2}\sigma(z)}\right)\right]
	\,, 
\eeq
where ${\rm erfc}(x)$ is the complementary error function defined by
\beq 
	{\rm erfc}(x)\equiv \frac{2}{\sqrt{\pi}}\int_x^{\infty}dz\exp(-z^2) \,. 
\eeq
For simplicity, we adopt the scaling relation for the photo-z error:
\beq
	\sigma(z) = 0.03\,(1+z) \,. 
\eeq

Apart from the calibration systematics for shear estimation, the main noise source for cosmic shear measurement is the intrinsic ellipticity of galaxies, which can be described as
\beq 
	N_\nu^{\gamma_i\gamma_j} = \delta_{ij}\frac{\ave{\gamma\rom{int}^2}}{\hat{n}_i} \,. 
\eeq
The quantities $\ave{\gamma\rom{int}^2}^{1/2}$ and $\hat{n}_i$ are the RMS intrinsic shear and the number density of galaxies per steradians in the $i$-th bin, respectively. We adopt the empirically derived value, $\ave{\gamma\rom{int}^2}^{1/2}=0.4$ \cite{BJ02}. Note that the quantity $\hat{n}_i$ is related to the average number density per arcminute square in the $i$-th bin, $\bar{n}_i$: 
\beq
	\hat{n}_i = 3600 \,\bar{n}_i\,\left(\frac{180}{\pi}\right)^2 \quad{\rm str}^{-1} \,. 
\eeq

Finally, in computing the Fisher matrix, Eq.(\ref{FisMat}), we replace the derivatives of the angular power spectrum $C_\l$ with respect to the cosmological parameters $p_i$ with a finite difference given by 
\beq
	\PD{C_\l}{p_i}(\vec{p})\simeq \frac{C_\l(p_i+\Delta p_i)-C_\l(p_i-\Delta p_i)}{2\Delta p_i}
	\,. \label{2side} 
\eeq
Specific values of the difference $\Delta p_i$ used in the analysis are listed in Table \ref{deriv-cl}.

\begin{table}
\bc
\begin{tabular}{ccccccccc} \hline 
$\Delta\Ob h^2$ & $\Delta\Om h^2$ & $\Delta\OL$ & $\Delta n\rom{s}$ & $\Delta A\rom{s}$ & $\Delta\tau$ & $\Delta\mnu$ & $\Delta w_0$ & $\Delta w_a$ \\ \hline 
0.05$\Ob h^2$ & 0.05$\Om h^2$ & 0.1$\Om$ & 0.005$n\rom{s}$ & 0.01$A\rom{s}$ & 0.4$\tau$ & 0.1$\mnu$ eV & 0.1 & 0.1 \\ \hline 
\end{tabular}
\ec
\vs{-1}
\caption{
Specific values of $\Delta p_i$ used to evaluate the derivatives of angular power spectra, Eq.~\eqref{2side}.
}
\label{deriv-cl}
\end{table}

\section{Results} \label{sec4}

In this section, forecast results for power spectrum measurements and parameter estimations are presented based on the canonical setup in previous section. In Sec.~\ref{sec4.1}, we first compute the signal-to-noise ratio for CMB lensing and cosmic shear observations. We then present the results of Fisher analysis for parameter forecast in Sec.~\ref{sec4.2}, just focusing on total neutrino mass ($\mnu$) and dark energy equation-of-state parameters ($w_0$ and $w_a$). In Sec.~\ref{sec4.3}, a role of cross correlations on the parameter constraints is discussed in some details. 

\subsection{Signal-to-noise ratio} \label{sec4.1}

\begin{figure}
\includegraphics[width=145mm]{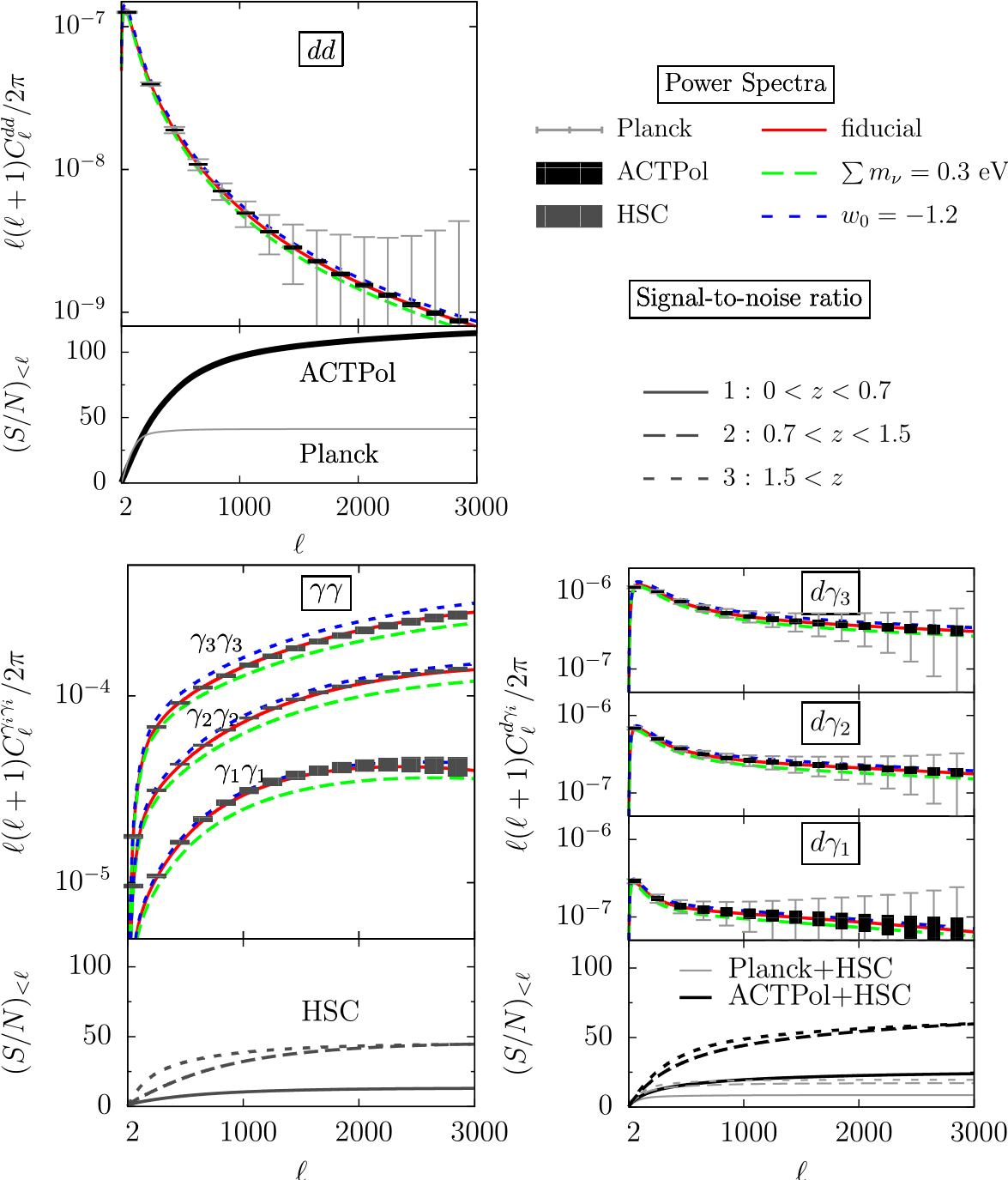}
\caption{
Angular power spectra for the deflection angle $C_\l^{dd}$ (top left), cosmic shear $C_\l^{\gamma_i\gamma_j}$ (bottom left) and deflection-shear cross correlations $C_\l^{d\gamma_i}$ (bottom right). The expected error bars are estimated from the canonical survey parameters with $\Delta\l=200$. For cosmic shear survey, the observed galaxies are divided into three subsamples; $0<z<0.7$, $0.7<z<1.5$ and $1.5<z$. Also we show the signal-to-noise ratio $(S/N)_{<\l}$ defined by Eq.~\eqref{S/N} in the bottom of each panel. 
}
\label{ClNl}
\end{figure}

To see how robustly the lensing power spectra can be measured with a high significance at each scale, let us first look at the expected signals and noises from the CMB lensing and cosmic shear measurements. Fig.~\ref{ClNl} plots the power spectra for CMB lensing (top left), cosmic shear (bottom left), and their cross correlation (bottom right) for fiducial cosmological model in Table \ref{Fid}. Also, we plot the variants of power spectra with slightly different values for $w_0$ and $\mnu$. The plotted errors for Planck, ACTPol, and HSC are estimated from 
\beq
	\Delta C_\l^{XY} = \begin{cases} 
		\displaystyle{\frac{C_\l^{XX}+N_\l^{XX}}{\sqrt{(\l+1/2)f\rom{sky}\Delta\l}}} & (X=Y) 
	\\ \\
	\displaystyle{\sqrt{\frac{(C_\l^{XY})^2+(C_\l^{XX}+N_\l^{XX})(C_\l^{YY}
		+ N_\l^{YY})}{(2\l+1)f\rom{sky}\Delta\l}}} & (X\not=Y) 
	\end{cases} 
	\,,
\eeq
where $f\rom{sky}$ is the sky coverage of each experiment, and $\Delta\l$ is the size of multipole bin, for which we set $\Delta\l=200$. In Figure \ref{ClNl}, we also show the signal-to-noise ratio for each power spectrum given by 
\beq 
	\left(S/N\right)_{<\l} = \sqrt{\sum_{\l'=2}^\l\left(\frac{C_{\l'}^{XY}}{\Delta C_{\l'}^{XY}}\right)^2}
	\,. \label{S/N}
\eeq
Note here that we set $\Delta\l=1$ to evaluate $\Delta C_{\l'}^{XY}$. 

In general, cosmic shear signal becomes larger and has a higher signal-to-noise ratio at smaller angular scales, and for the higher redshift sources. Since the free-streaming suppression of the massive neutrinos is known to appear at relatively smaller scales (or higher multipoles), the cosmic shear signals are generally sensitive to the change of the total mass of neutrinos, as well as to dark energy properties through the late-time variation of structure growth. Figure \ref{ClNl} indicates that the cosmic shear signals from HSC survey potentially have enough sensitivity to detect total neutrino mass of $\sim0.1$ eV and to constrain dark energy properties. 

By contrast, the angular power spectrum of deflection angles from CMB lensing differs from that of the cosmic shear, and has a larger amplitude at lower multipoles, as mentioned in Sec.~\ref{sec2}. Thus, the CMB lensing seems less sensitive to the suppression effect of massive neutrinos, and a high-angular resolution experiment is required for measuring the neutrino masses with a sub-eV. As shown in Fig.~\ref{ClNl}, with ACTPol, we can clearly discriminate between the total neutrino masses of the difference $\sim0.2$ eV from the CMB lensing experiment alone. Note that the signal-to-noise ratio for CMB lensing from the ACTPol is rather higher than that of the cosmic shear signal obtained from the HSC survey. We thus naively expect that the ACTPol can give a better constraint on neutrino mass than the HSC. In practice, however, there are a sizable amount of degeneracies among several cosmological parameters. In particular, as indicated by Fig.~\ref{deriv}, the neutrino mass is tightly correlated with $\Om h^2$. Thus, the final outcome of the neutrino mass constraint, after marginalizing over the other parameters, cannot be straightforwardly understood only from the signal-to-noise ratios. 

Finally, we note that the cross correlation signal $d\gamma_i$ obtained from ACTPol and HSC has a high signal-to-noise ratio (bottom-right panel of Fig.~\ref{ClNl}). We also compute the signal-to-noise ratios for other cross correlation signals, and find that temperature-deflection cross correlation $\Theta d$ from ACTPol and Planck would be measured with high signal-to-noise ratio, S/N$\sim \mcO(10)$, while the temperature-shear cross correlations $\Theta\gamma_i$ have a low signal-to-noise ratio, S/N$\lesssim1$. These results basically come from the facts that the non-vanishing cross correlations can be attributed to the ISW effect, and a large signal of ISW effect only appear at large-angular scales. 

\subsection{Parameter forecast}\label{sec4.2}

We now present the forecast results for cosmological parameters derived from Planck, ACTPol and HSC. Table \ref{Fis} summarizes the marginalized $1\sigma$ (68\%) errors for the results combining two (Planck and HSC) or three observations (Planck, ACTPol and HSC). In each case, we examine following three cases: 
\bi
\item $+d$ : including CMB lensing data ($C_\l^{dd}$ and $C_\l^{\Theta d}$) 
\item $+\gamma$ : including cosmic shear data ($C_\l^{\gamma_i\gamma_j}$ and $C_\l^{\Theta\gamma_i}$)
\item $+d+\gamma$ : combining all power spectra, i.e., $C_\l^{dd}$, $C_\l^{\Theta d}$, $C_\l^{\gamma_i\gamma_j}$, $C_\l^{\Theta\gamma_i}$, and $C_\l^{d\gamma_i}$ 
\ei
Note that we use the lensing information up to $\l\rom{max}=3000$, and add the primary (unlensed) CMB data (i.e., $C^{\Theta\Theta}_\l$, $C^{\Theta E}_\l$ and $C^{EE}_\l$) as prior information in all cases. We assume that the observed area of HSC is entirely overlapped with that of ACTPol, and similarly the survey region of ACTPol is totally included in the nearly full-sky survey with Planck. 

Overall, the constraints obtained from the single experiments ($+d$ or $+\gamma$) are almost comparable, and combining all the lensing observations ($+d+\gamma$) moderately improves the constraints on cosmological parameters. The constraints on $w_0$ and $w_a$ obtained from the Planck and HSC roughly correspond to those from the Stage-III class experiments defined by Dark Energy Task Force \cite{Alb06} (e.g., the Dark Energy Survey and Pan-STARRS-4), and the constraints are further improved by a factor of $\sim2$ when we add the ACTPol data. The neutrino mass constraint is also improved by a factor of $\sim1.5$, and with Planck, ACTPol and HSC, the total mass of neutrinos can be detected with $\sim2\sigma$ significance for a fiducial value $\mnu=0.1$eV. For other cosmological parameters, the constraints by adding ACTPol data are $1.5$-$2$ times better than those obtained from Planck and HSC. Note that our neutrino mass constraint from the CMB lensing is roughly consistent with those obtained by Refs.~\cite{LPPP06,dP09,Niemack:2010wz,Kaplinghat:2003bh}. Also, the constraints on $w_0$ and $w_a$ from cosmic shear measurement roughly match the results obtained in Ref.~\cite{HTW06}. In other words, including the neutrino mass in the parameter estimation would not drastically alter the constraints on dark energy equation-of-state parameters from the lensing measurements. 

\begin{figure}
\bc 
\includegraphics[width=14cm,clip]{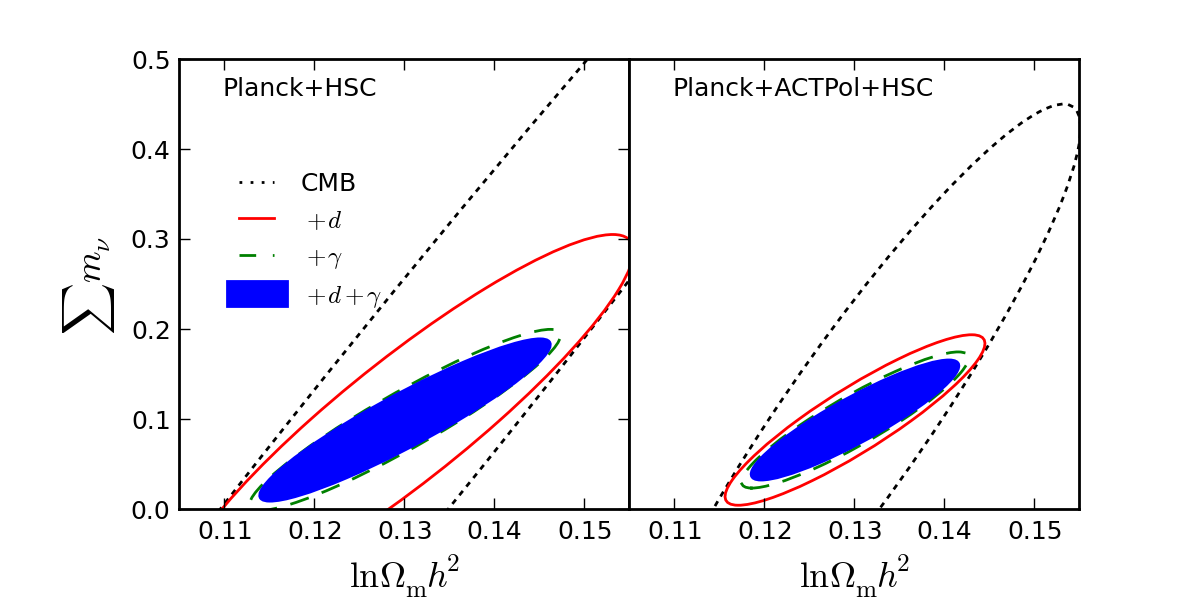} 
\ec 
\caption{
Expected $1\sigma$ error contours on $\ln\Om h^2$-$\mnu$ plane assuming Planck and HSC (left), and Planck, ACTPol and HSC (right). The gray dotted lines represent the results from primary CMB, while the red solid and green dashed lines indicate the results further including CMB lensing and cosmic shear, respectively. The blue filled ellipse are the results of all measurements.
}
\label{Omh-mnu}
\end{figure}

\begin{figure}
\bc
\includegraphics[width=14cm,clip]{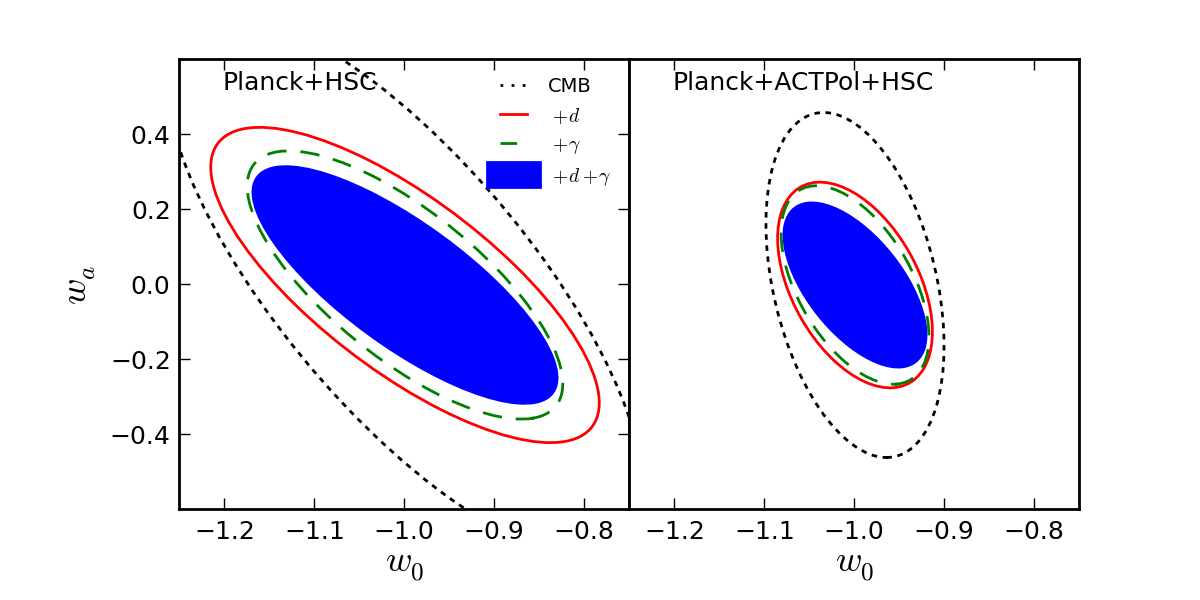}
\ec 
\caption{
Same as Fig.~\ref{Omh-mnu} but for expected $1\sigma$ error contours on $w_0$-$w_a$ plane.
}
\label{w0-wa}
\end{figure}

Figs.~\ref{Omh-mnu} and \ref{w0-wa} elucidate the impact of upcoming lensing experiments on neutrino mass and dark energy properties by showing the expected $1\sigma$ ($68\%$C.L.) contours on $\ln\Om h^2$ and $\mnu$, and $w_0$ and $w_a$, respectively. Left panels are the results obtained from the Planck and HSC survey, while the right panels are obtained by further adding the ACTPol data. Comparing with the constraints coming from the primary CMB data alone, the size of error ellipses becomes rather reduced when we add the lensing information, but the relative differences between the three cases (i.e., $+d$, $+\gamma$ and $+d+\gamma$) are basically small, as we mentioned above. Further, the inclinations of the ellipses are almost the same. This can be partly explained by the fact that cosmological information on dark energy and massive neutrinos mostly comes from lower redshifts ($z<1$), and thus, the cosmic shear and CMB lensing power spectra have almost identical information on the dark energy properties and neutrino mass, though the CMB lensing signal has potentially sensitive to higher redshifts. As shown in Fig.~\ref{ClNl}, there exists non-vanishing correlation $d\gamma_i$, and thereby the cosmic shear and CMB lensing signals cannot be regarded as statistically independent signals. As a result, only a moderate improvement of the constraints on the parameters, $\mnu$, $w_0$ and $w_a$, is achieved, and a tight correlation between cosmological parameters, $\mnu$ and $\ln\Om h^2$, $w_0$ and $w_a$, still remains in the combined results of two lensing experiments. Nevertheless, the CMB lensing and cosmic shear experiments greatly improve the constraints from the primary CMB information, and these can be used as an independent cross check for extracting cosmological information in an unbiased way. Hence, science benefit for combining two lensing experiments is still valuable.

\begin{table}
\bc 
\begin{tabular}{ccccccc} \hline & \multicolumn{3}{c}{Planck+HSC} & 
\multicolumn{3}{c}{Planck+ACTPol+HSC} \\ 
parameter & $+d$ & $+\gamma$ & $+d+\gamma$ & $+d$ & $+\gamma$ & $+d+\gamma$ \\ \hline 
$\ln(\Ob h^2)$ & 0.0063 & 0.0058 & 0.0057 & 0.0030 & 0.0030 & 0.0029 \\ 
$\ln(\Om h^2)$ & 0.016  & 0.011  & 0.010  & 0.0094 & 0.0080 & 0.0074 \\ 
$\OL$          & 0.031  & 0.022  & 0.021  & 0.017  & 0.014  & 0.013  \\ 
$w_0$          & 0.14   & 0.12   & 0.11   & 0.056  & 0.054  & 0.052  \\ 
$w_a$          & 0.28   & 0.24   & 0.21   & 0.18   & 0.17   & 0.14   \\ 
$n\rom{s}$     & 0.0033 & 0.0029 & 0.0028 & 0.0023 & 0.0023 & 0.0022 \\ 
$\ln(A\rom{s})$ & 0.014  & 0.012  & 0.012  & 0.011  & 0.011  & 0.010  \\ 
$\tau$         & 0.0040 & 0.0040 & 0.0040 & 0.0040 & 0.0040 & 0.0040 \\ 
$\mnu$ [eV]    & 0.13   & 0.060  & 0.055  & 0.061  & 0.046  & 0.041  \\ \hline 
\end{tabular}
\ec
\vs{-1}
\caption{
Forecast results for marginalized $1\sigma$ errors for each cosmological parameter, assuming Planck and HSC (left three columns) and Planck, HSC and ACTPol (right three columns). The labels, $+d$, $+\gamma$, and $+d+\gamma$, respectively indicate the cases including CMB lensing, cosmic shear, and combining both. The primary CMB information is included in all cases. 
}
\label{Fis} 
\end{table}

\subsection{Role of cross correlation statistics} \label{sec4.3}

Here, we discuss the role of the cross correlation statistics to the parameter constraints given in Table \ref{Fis}. Apart from the CMB temperature-polarization cross correlation $\Theta E$ and tomographic lensing correlations $\gamma_i\gamma_j$, there are three kinds of measurable cross correlations, i.e., $\Theta d$, $\Theta\gamma_i$, and $d\gamma_i$. Among these, the deflection-shear cross correlations have almost identical information to the deflection or shear auto correlation, and they do not significantly contribute to the neutrino and dark energy equation-of-state parameters, as we already discussed in previous section. Also, the temperature-shear cross correlation has a small signal-to-noise ratio, and no valuable cosmological information can be obtained. 

On the other hand, the signal-to-noise ratio for temperature-deflection cross correlation is S/N$\sim \mcO({10})$ for the Planck and ACTPol experiments. Since the non-vanishing contribution of this cross correlation basically comes from the ISW effect, it is expected to be sensitive to the early-time evolution of dark energy. Although this contribution has been previously ignored in the Fisher analysis in Ref.~\cite{dP09}, where they compute the constraints on $w_0$ and $w_a$ from the CMB lensing signal alone, we find that this signal gives an interesting contribution to the constraint on the time dependence of the dark energy equation-of-state parameter, $w_a$. 

\begin{figure}
\bc
\includegraphics[width=95mm]{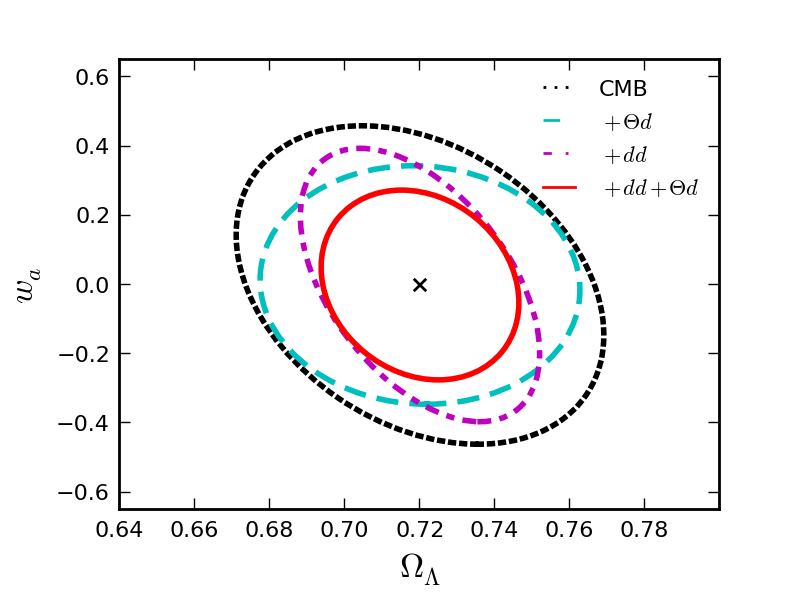}
\ec
\caption{
Expected $1\sigma$ error contours on $\OL$-$w_a$ plane. The dot-short-dashed line (magenta) show the constraints obtained from deflection auto correlation ($+dd$), and the dot-long-dashed line (cyan) is the result from temperature-deflection cross correlation ($+\Theta d$). The red solid line is the combined results from all signals of the CMB lensing. Note that primary CMB information is added in all three cases. As a reference, we also plot the result from primary CMB information (gray dotted).
}
\label{dISW}
\end{figure}

To see the significance of the contribution from temperature-deflection cross correlation, in Fig.~\ref{dISW}, we plot the two-dimensional contour of $1\sigma$ error on $w_a$-$\OL$ plane, assuming the CMB lensing data from Planck and ACTPol. Here, in addition to the constraints coming from the primary CMB information (dashed, labeled as CMB), we consider the following three cases: 
\bi
\item $+dd$ : using $C_\l^{dd}$,
\item $+\Theta d$ : using $C_\l^{\Theta d}$,
\item $+\Theta d+dd$ : using $C_\l^{\Theta d}$, $C_\l^{dd}$ 
\ei
In all three cases, we add the primary CMB information ($C_\l^{\Theta\Theta}$, $C_\l^{\Theta E}$, and $C_\l^{EE}$), and in computing the Fisher matrix for $+dd$ and $+\Theta d$ cases, the contributions coming from the CMB lensing auto/cross correlation are separately computed and are then added to the Fisher matrix for primary CMB information. Note here that the $+\Theta d+dd$ case just coincides with the $+d$ case examined in previous section. 

As we see from Fig.~\ref{dISW}, the temperature-deflection cross correlation can give a slightly better constraint than that from the deflection auto correlation. This is somewhat remarkable in the sense that the signal-to-noise ratio for temperature-deflection cross correlation is rather lower than that of the deflection auto correlation. The reason for this is presumably explained by the fact that the inclusion of the cross correlation breaks the degeneracy between $\OL$ and $w_a$: the correlation coefficient between $\OL$ and $w_a$, given by $r(\OL,\,w_a)=\{\bm{F}^{-1}\}_{ij}/[\{\bm{F}^{-1}\}_{ii}\{\bm{F}^{-1}\}_{jj}]^{-1/2}$ with $i=w_a$ and $j=\OL$, changes the value from $r(\OL,\,w_a)=0.3$ to $-0.1$, relatively reducing the statistical correlation between these parameters. The result indicates that the temperature-deflection cross correlation has a better sensitivity to higher redshifts compared to the auto correlation, and potentially gives a tight constraint on the time variation of dark energy equation-of-state parameter, $w(z)$. In appendix \ref{appA}, this point will be studied in detail based on the principal component analysis. 

\section{Sensitivity to the survey design of cosmic shear experiment} \label{sec5}

In this section, to elucidate the robustness or sensitivity of the forecast results in previous section, we allow to vary the setup of cosmic shear experiment, while we keep adopting the Subaru HSC facility. Then, we examine how the constraints on the neutrino mass and dark energy properties are changed. In Sec.~\ref{sec5.1}, we first discuss how to characterize several key parameters of the cosmic shear experiment as a function of observation time or exposure time. In Sec.~\ref{sec5.2}, the forecast results for the constraints on neutrino mass and dark energy properties are given, and the sensitivity to the observation time or exposure time is investigated. 

\subsection{Modeling cosmic shear surveys} \label{sec5.1}

In a cosmic shear survey, the exposure time, total observation time, and number of redshift bins are essential key parameters to determine a survey design, and these are related to the sky coverage, survey depth and number of galaxy samples. Thus, the final outcome of the parameter constraints can be sensitively affected by those parameters. Here, we wish to relate the parameters of survey design, and try to characterize the forecast results for neutrino mass and dark energy equation-of-state parameters as functions of the most important parameters, i.e., total observation time and exposure time. To do this, we basically follow the treatment by Ref.~\cite{YPHNS07}. That is, for a given number of redshift bin $N$, we relate the sky coverage $f\rom{sky}$, mean redshift $z_m$, the total number density of galaxies per square arcminute $N_g$ introduced in Sec.~\ref{sec3} to the exposure time $t\rom{exp}$ and total observation time $T\rom{obs}$. Supposing the Subaru HSC facility for a cosmic shear experiment, we then express the forecast constraints as functions of $T\rom{obs}$ and $t\rom{exp}$. 

In the tomographic lensing technique, several filters are used to construct the redshift subsamples, and the available number of redshift bins is basically determined by the number of filters. Thus, we must first specify the number of filter and determine the exposure time for each filter. Following Ref.~\cite{YPHNS07}, we assume that the $i$-band filter is used for the case $N=1$, and the $i$- and $r$-band filters are used for $N=2$. As for the cases with $N=3$ and $4$, we consider $g$-, $r$-, $i$- and $z$-band observations. Based on the survey proposal in Ref.~\cite{HSC}, for the one field-of-view observation, the exposure time of the $r$- and $g$-bands is assumed to be equal to that of the $i$-band filter, but the $z$-band exposure time is set to be $4/3$ times longer. Then, depending on the number of filters or redshift bins, the total exposure time $\sum t\rom{exp}$ per one field-of-view is effectively expressed as the $i$-band exposure time, which we denote by $t\rom{exp}$. The relation between $\sum t\rom{exp}$ and $t\rom{exp}$ is summarized in Table \ref{tom}. 

We now relate the parameters $z_m$, $N_g$ and $f\rom{sky}$ to the exposure time $t\rom{exp}$ and total observation time $T\rom{obs}$. The mean redshift $z_m$ for the observed galaxy distribution depends on the survey depth, and is related to the exposure time per one field-of-view. According to Ref.~\cite{YPHNS07}, a simple scaling relation between $z_m$ and $t\rom{exp}$ is given by 
\beq
	z_m(t\rom{exp}) = z_{m,0}\left(\frac{t\rom{exp}}{t_0}\right)^{0.067}  \,, \label{eq:z_m}
\eeq
where we set $t_0=15$min and $z_{m,0}=1$\footnote{Ref.~\cite{YPHNS07} adopted $t_0\sim 30$ min and $z_{m,0}=0.9$ to just follow the parameters described in \cite{AR07}. Here, we adopt different values to match the recent study on the HSC survey plan \cite{HSC}.}. As for the total number of galaxies per square arcminute $N_g$, we adopt the empirical relation determined by the pilot observation \cite{HSC}: 
\beq
	N_g(t\rom{exp}) = 35\times\left(\frac{t\rom{exp}}{t_0}\right)^{0.3}{\rm arcmin}^{-2}
	\,. \label{eq:N_g}
\eeq
Finally, adopting the 1.8 deg$^2$ of the field-of-view of HSC \cite{HSC}, the sky coverage $f\rom{sky}$ is expressed as the function of $t\rom{exp}$ and total observation time $T\rom{obs}$. According to Ref.~\cite{YPHNS07}, we have 
\beq 
	f_{\rm sky}(t_{\rm exp}) = 1.8\,{\rm deg}^2\,\frac{T_{\rm obs}}{1.1\sum t_{\rm exp}+t_{\rm ove}}
		\times \frac{0.05}{2,000 \,{\rm deg}^2}
	\,, \label{eq:f_sky}
\eeq
Here, we take into account the overhead time $t\rom{ove}=5$ min and the processing time $0.1\times\sum t\rom{exp}$. 

\begin{figure}
\bc
\includegraphics[width=95mm,clip]{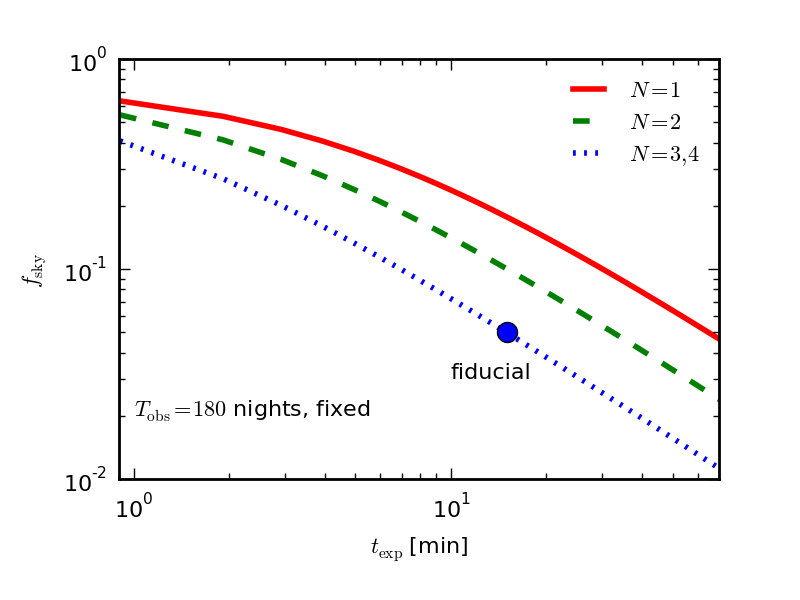}  \vs{-2}
\ec
\caption{
Sky coverage $f\rom{sky}$ of cosmic shear survey as a function of $i$-band exposure time $t\rom{exp}$ for different number of redshift bins; $N=1$ (red, solid), $N=2$ (green, long-dashed) and $N=3,\,4$ (blue, short-dashed). Here, we fix the total observation time $T\rom{obs}$ to $180$\,nights. The blue point on this line is located at $T\rom{obs}=180$ nights where the survey parameters correspond to the values used in Sec.~\ref{sec4}.
} 
\label{fsky-texp}
\end{figure}

To sum up, given the exposure time $t\rom{exp}$ and total observation time $T\rom{obs}$, the redshift distribution of galaxies characterized by Eq.~\eqref{eq:ngal_z} is specified and the sky coverage of cosmic shear survey is fixed through the relations \eqref{eq:z_m}-\eqref{eq:f_sky}, depending on the number of filters or redshift bins (see Table \ref{tom}). Then, we can proceed to the Fisher analysis just following the procedure in Sec.~\ref{sec3}. Finally, we note that the canonical setup for the cosmic shear experiment examined in Sec.~\ref{sec4} corresponds to the description mentioned above with the specific values of $t\rom{exp}=15$ min, $T\rom{obs}=180$ nights\footnote{We assume 8 hours per night throughout the analysis.} and $N=3$. In Fig.~\ref{fsky-texp}, as a reference, the dependence of the sky coverage $f\rom{sky}$ on the number of redshift bins $N$ is plotted against the $i$-band exposure time $t\rom{exp}$, fixing the observation time to $T\rom{obs}=180$ nights. 

\begin{table}
\bc
\begin{tabular}{ccccc} \hline 
$N$ & 1 & 2 & 3 & 4 \\ \hline 
Filter & $i$ & $i$ + $r$ & $i$ + $r$ + $g$ + $z$ & $i$ + $r$ + $g$ + $z$ \\ \hline 
$\sum t\rom{exp}$ [min] & $t\rom{exp}$ & $2t\rom{exp}$ & $(65/15)t\rom{exp}$ & $(65/15)t\rom{exp}$ \\ \hline 
Redshift bins & All redshift range & $z<z_m$ & $z<0.7$ & $z<0.5$ \\ 
 & & $z_m<z$ & $0.7<z<1.5$ & $0.5<z<z_m$ \\
 & & & $1.5<z$ & $z_m<1.5$ \\
 & & & & $1.5<z$ \\ \hline 
\end{tabular}
\ec
\vs{-1}
\caption{
The number of redshift bin, $N$, redshift range for each redshift bin and $\sum t\rom{exp}$ for tomography. We use the $i$-band filter for the case $N=1$, and the $r$- and $i$-band filters for $N=2$. For the cases with $N=3$ and $4$, we consider $g$-, $r$-, $i$- and $z$-band observations.
}
\label{tom}
\end{table}

\subsection{Results} \label{sec5.2}

Based on the procedure in the previous subsection, we vary the setup of cosmic shear survey, and derive the constraints on the neutrino mass and dark energy equation-of-state parameters, characterized by the exposure time and total observation time. In what follows, combining the CMB lensing and primary CMB information obtained from Planck and ACTPol, the results are separately presented for neutrino mass (Sec.~\ref{sec5.2.1}) and dark energy properties (Sec.~\ref{sec5.2.2}). 

\subsubsection{Neutrino mass} \label{sec5.2.1}

\begin{figure}
\bc
\includegraphics[width=10cm]{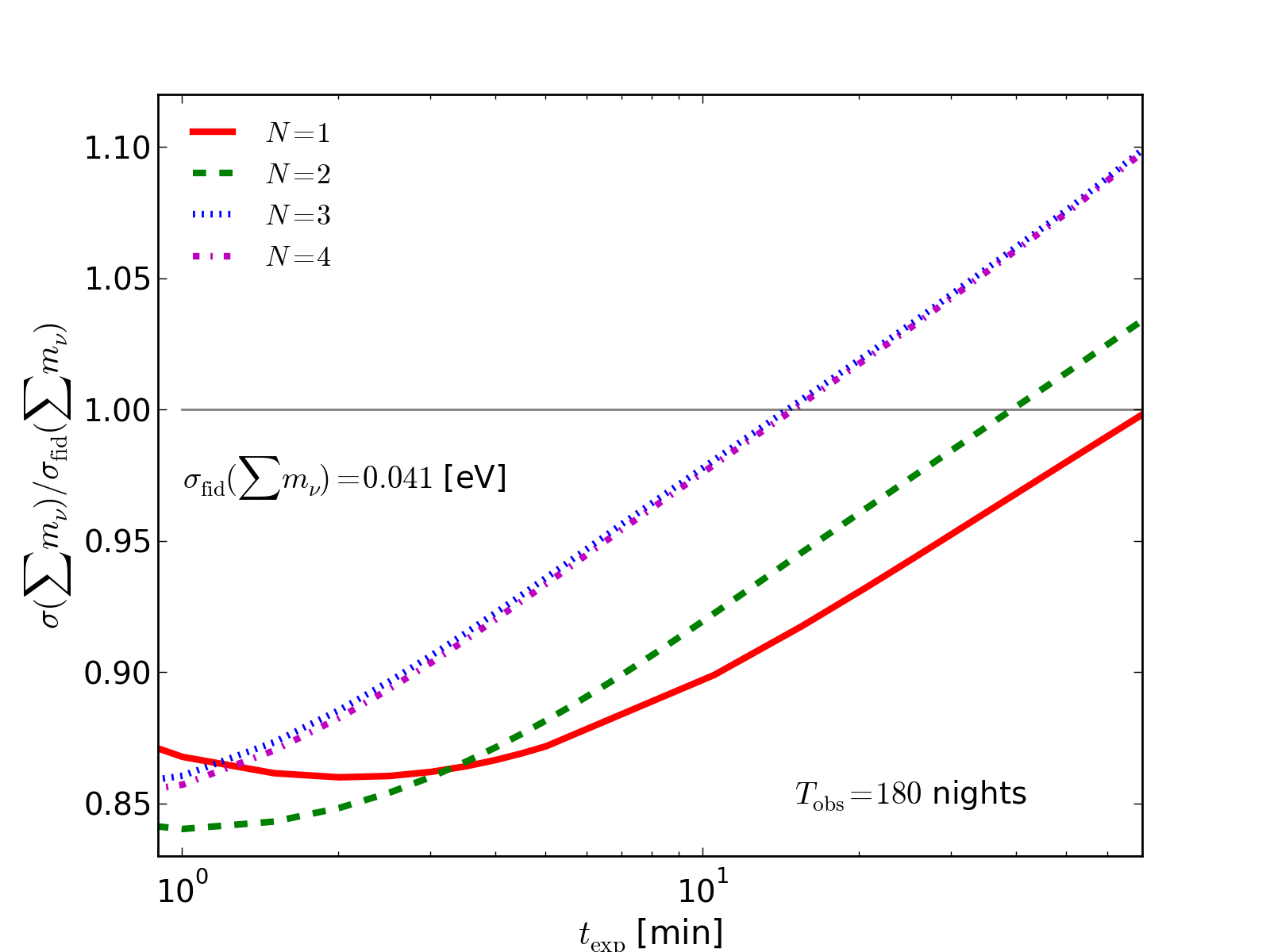}  \vs{-1.5}
\ec
\caption{
The constraint on total neutrino mass, $\sigma(\mnu)$, as a function of the exposure time $t\rom{exp}$ for $N=1$ (red solid), $N=2$ (green long-dashed), $N=3$ (blue short-dashed) and $N=4$ (magenta dotted). The total observation time $T\rom{obs}$ is fixed to $180$ nights. The resultant value of $\sigma(\mnu)$ is plotted dividing by the value obtained in Sec.~\ref{sec4}, $\sigma\rom{fid}(\mnu)$.
} 
\label{mnu-texp}
\end{figure}

Let us first show the forecast results for the constraint on the total neutrino mass, fixing the total observation time to $T\rom{obs}=180$ nights. In Fig.~\ref{mnu-texp}, we plot the marginalized $1\sigma$ error, $\sigma(\mnu)$, as a function of the $i$-band exposure time $t\rom{exp}$. Here, the results are normalized by the value obtained from the canonical setup in Sec.~\ref{sec4}, i.e., $\sigma\rom{fid}(\mnu)=0.041$ eV. Note again that all the power spectra are computed up to $\l\rom{max}=3000$ to calculate the Fisher matrix. 

Basically, the resultant marginalized $1\sigma$ error is shown to be a monotonically increasing function of the exposure time. That is, the neutrino mass constraint is improved as decreasing the exposure time, and at $t\rom{exp}\gtrsim3$\,min, the tightest constraint is obtained from the $N=1$ case. This indicates that for a limited observation time, a shallow and wide-field survey is preferable, and no tomographic technique is necessary to improve the neutrino mass constraint. This result is contrasted with Ref.~\cite{HTW06}, where they fixed the exposure time and sky coverage, and showed that the constraint on neutrino mass is improved with the number of redshift bins. However, the resultant survey setup for lensing tomography requires an increasingly long observation time, which seems in some sense impractical.

A closer look at a shorter exposure time at $t\lesssim3$\,min reveals that the constraint with $N=1$ ceases to improve and turn next to gradually worsen. Eventually, the $N=2$ case can give a better constraint on the neutrino mass. The basic reason for this behavior comes from the competition of the two effects. That is, as decreasing the exposure time, the survey area of cosmic shear experiment increases and the statistical error is reduced. However, as a trade off, the redshift distribution of galaxy samples becomes shallower and the number of samples per square arcminute tends to decrease, leading to the reduction of the signal-to-noise ratio. Hence, for $N=1$ case, there appears an optimal exposure time, $t\rom{exp}\simeq3$ min, where the tightest constraint on the neutrino mass is obtained.  

Note that the optimal value of $t\rom{exp}$ does also exist for $N\ne1$ cases, and these would be much shorter than that of the $N=1$ case. In such a short exposure time with $t\rom{exp}\lesssim5$ min, however, a robust and reliable shear measurement is practically difficult and challenging due to the stability of the seeing condition. Thus, we hereafter consider the exposure time of $t\rom{exp}\geq5$ min, and look for the best constraint on the neutrino mass, characterized by the function of total observation time. 

\begin{figure}
\bc
\includegraphics[width=10cm]{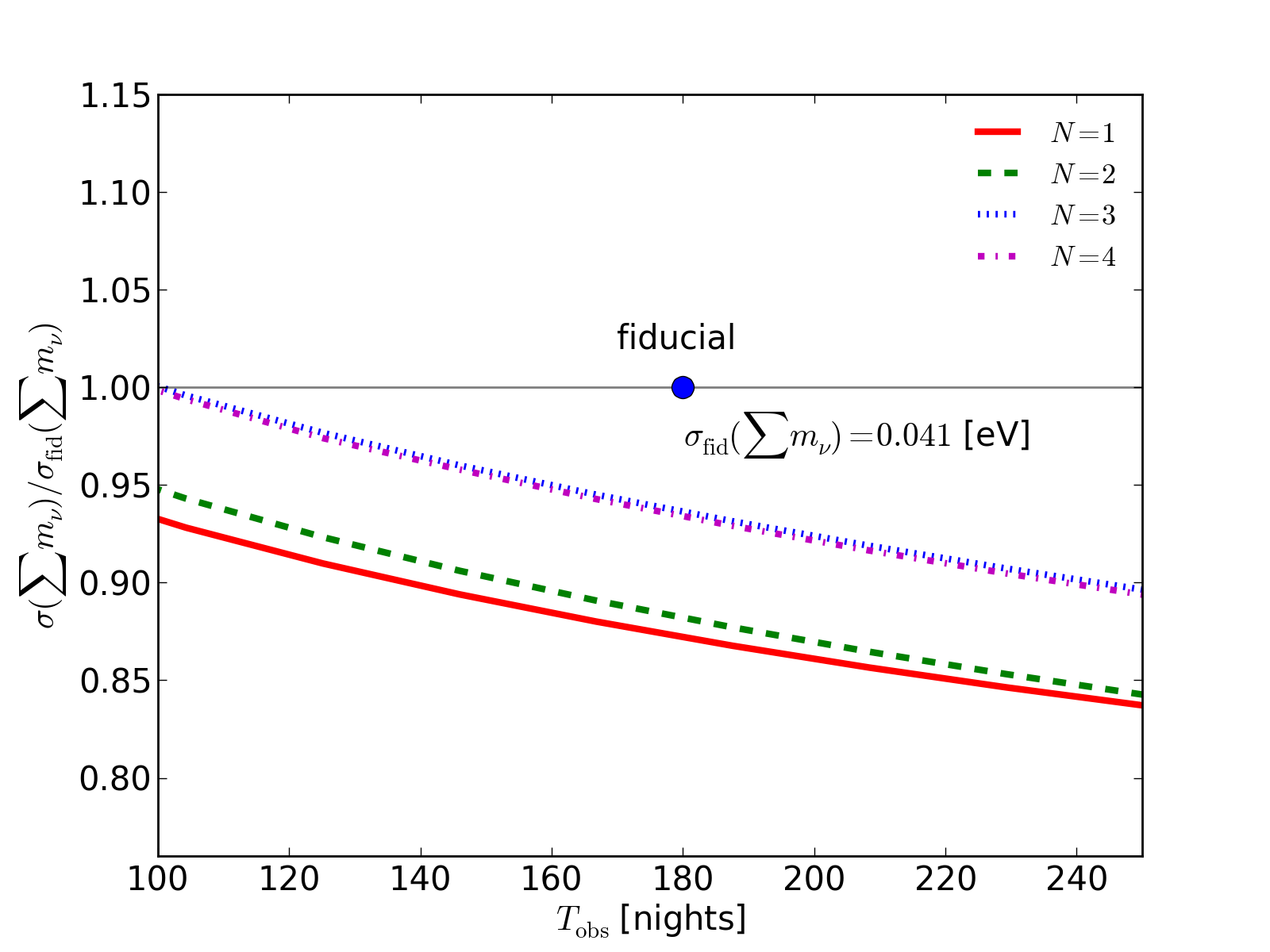}  \vs{-1.5}
\ec
\caption{
The expected neutrino mass constraint $\sigma(\mnu)$ as a function of the total observation time $T\rom{obs}$, determined by the optimal choice of the exposure time. Here, the exposure time is optimized so as to minimize $\sigma(\mnu)$ for each observation time under the condition $t\rom{exp}>5$ min. The blue filled circle is the result for the fiducial survey setup; $N=1$ (red solid), $N=2$ (green long-dashed), $N=3$ (blue short-dashed) and $N=4$ (magenta dotted).
}
\label{mnu-tobs}
\end{figure}

Fig.~\ref{mnu-tobs} shows the neutrino mass constraint plotted against the total observation time $T\rom{obs}$, which is determined by the optimal choice of the exposure time, so as to minimize the marginalized $1\sigma$ error $\sigma(\mnu)$ for each observation time, under the condition $t\rom{exp}\geq5$ min. Again, the results are normalized by the fiducial value, $\sigma\rom{fid}(\mnu)$, obtained in the previous section. As anticipated from Fig.~\ref{mnu-texp}, the optimal neutrino mass constraint is given as an decreasing function of $T\rom{obs}$, and roughly scales as $T\rom{obs}^{-0.5}$. Under the condition $t\rom{exp}\geq5$\,min, the $N=1$ case can give the tightest constraint over the plotted range of $T\rom{obs}$. However, the resultant improvement factor, given by the inverse of $\sigma(\mnu)/\sigma\rom{fid}(\mnu)$, is rather small, and it somehow reaches at $\sim1.2$ at $T\rom{obs}=250$\,nights. The dependence on the number of redshift bins is also very weak, and no significant improvement is expected from the optimization of the exposure time. 

\subsubsection{Dark energy} \label{sec5.2.2}

Next consider the constraints on dark energy properties, i.e., $w_0$ and $w_a$. In order to find the optimal exposure time for the dark energy equation-of-state parameters, we here introduce the figure-of-merit (FoM) defined by \cite{Alb06}
\beq 
	{\rm FoM}(w_0,w_a) = \frac{1}{\sqrt{\det\tilde{\bm{F}}^{-1}}}  \,, \label{FoM}
\eeq
where $\tilde{\bm{F}}^{-1}$ is the 2$\times$2 sub-matrix whose components are taken from the inverse of Fisher matrix associated with the parameters $w_0$ and $w_a$. Since the determinant of $\tilde{\bm{F}}^{-1}$ is proportional to the area of the two-dimensional error ellipse on $w_0$-$w_a$ plane, a larger FoM value implies a tighter constraint on dark energy properties. 

\begin{figure}
\bc
\includegraphics[width=10cm]{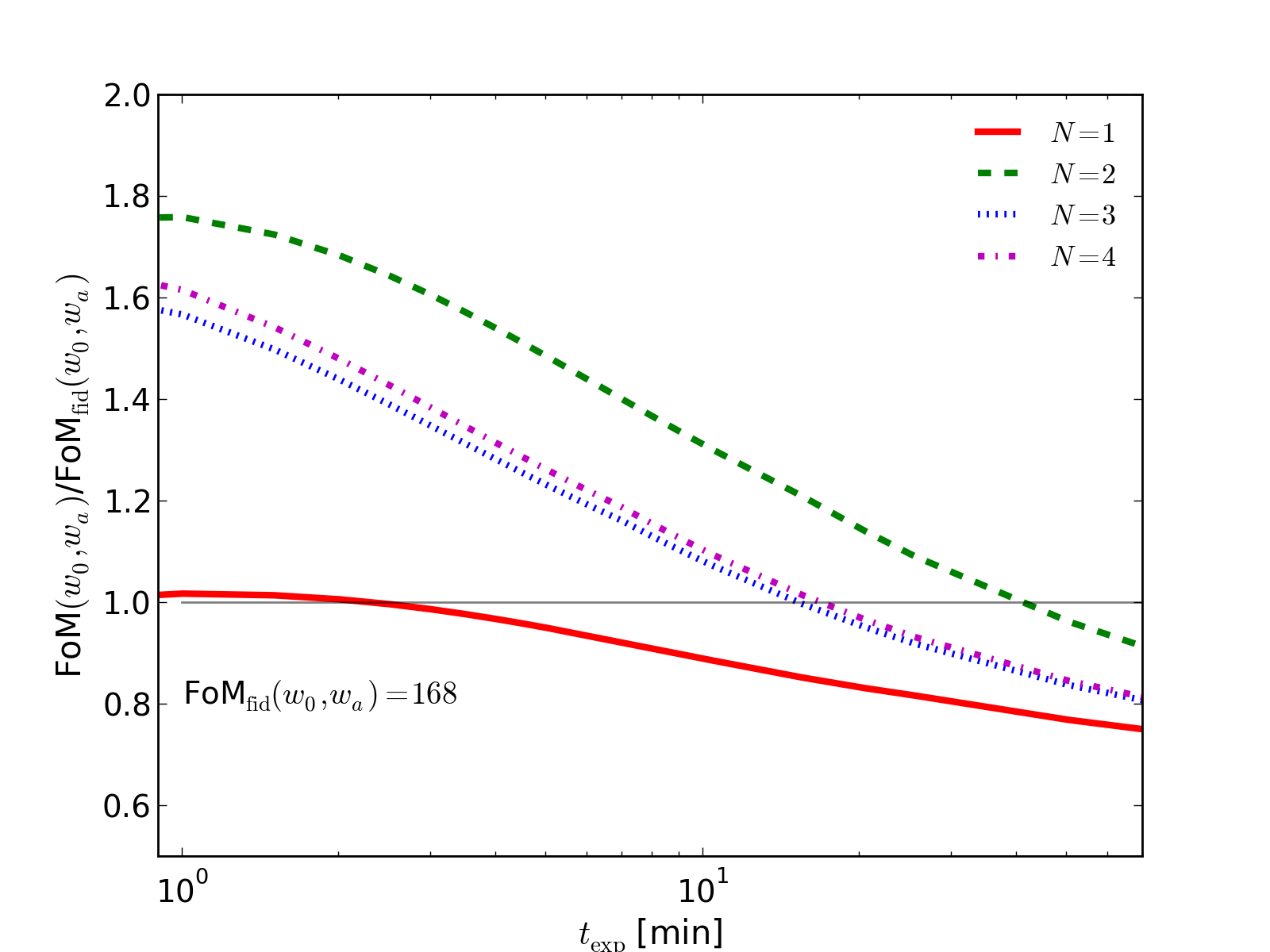}  \vs{-1.5}
\ec
\caption{
The figure-of-merit, FoM($w_0,w_a$), as a function of exposure time, fixing total observation $180$ nights. The results are normalized by FoM$\rom{fid}$($w_0,w_a$)$=168$, which is obtained from the canonical survey setup (see Sec.~\ref{sec4}). Line types are the same as in Fig.~\ref{mnu-texp}.
} 
\label{fom-texp}
\end{figure}

In Fig.~\ref{fom-texp}, fixing the total observation time to $T\rom{obs}=180$ nights, we show the FoM as the function of exposure time, and the results are normalized by the value obtained from the canonical setup in Sec.~\ref{sec4}, i.e., FoM$\rom{fid}(w_0,w_a)=168$. The FoM is a decreasing function of exposure time, and among several choices of tomographic binning, the $N=2$ case can give a better performance for the FoM. While this result is consistent with Ref.~\cite{YPHNS07}, the situation they considered is somewhat different from ours: they considered the cosmic shear experiment alone assuming the massless neutrinos. Also, the optimal number of filters is contrasted with the case of the neutrino mass constraint, i.e., $N=1$. Presumably, the differences between the optimal number of filters would reflect the fact that the dark energy properties sensitively affect the growth of the large-scale structure, while the effect of the massive neutrinos can give a free-streaming suppression on the small-scale structure growth, whose redshift dependence is extremely weak. In this sense, the tomographic technique of the cosmic shear experiment is especially helpful in constraining the time evolution of dark energy equation-of-state, and for our current setup with limited observation time, the choice of $N=2$ becomes optimal. Note that the choice of $N=2$ still remains optimal even if we look for the maximum value of FoM including the neutrino mass. 

\begin{figure}
\bc
\includegraphics[width=10cm]{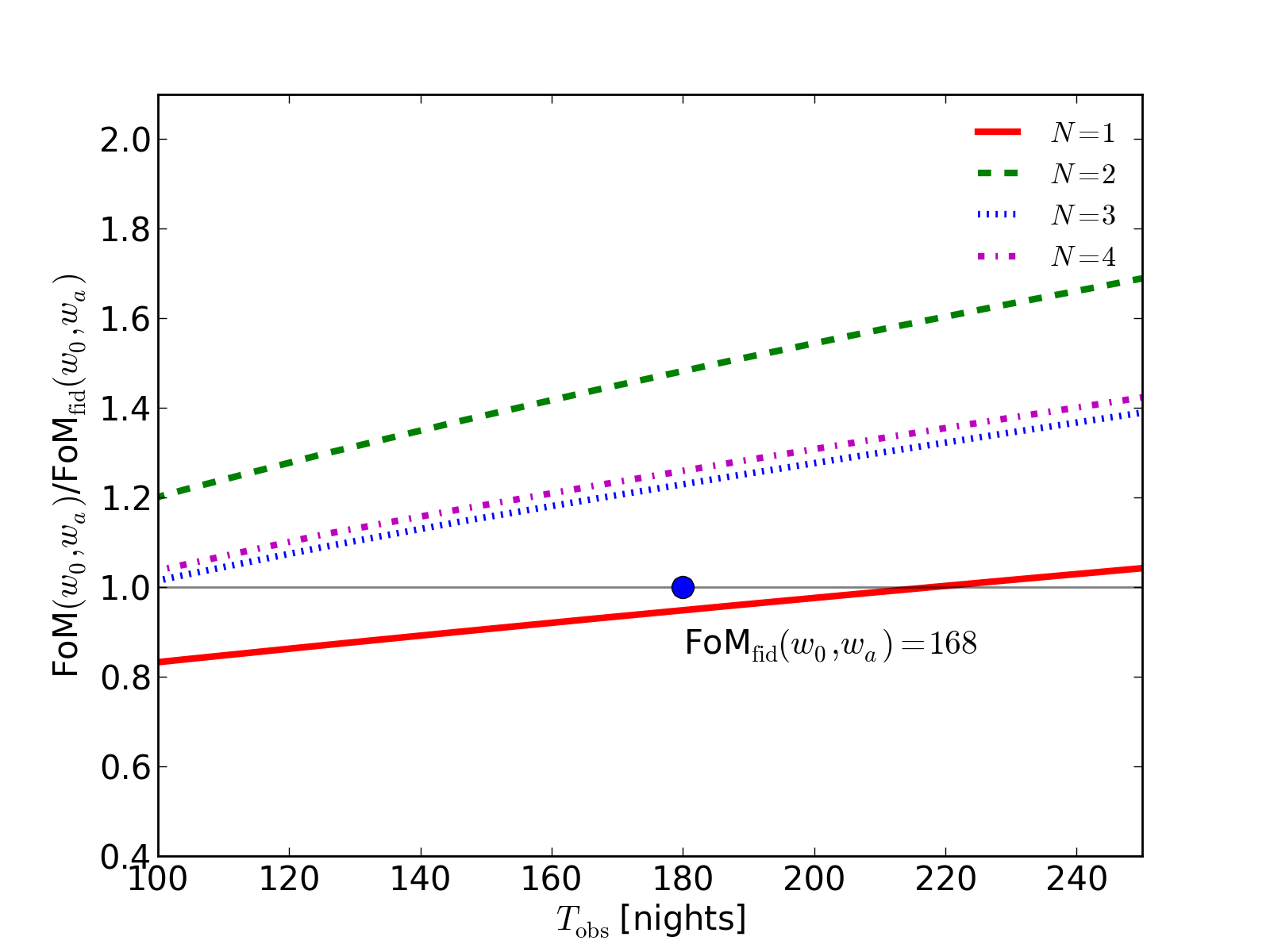}  \vs{-1.5}
\ec
\caption{
The figure-of-merit, FoM($w_0,w_a$), as a function of total observation time $T\rom{obs}$, determined by the optimal choice of the exposure time. Here, the exposure time is optimized so as to maximize FoM($w_0,w_a$) for each observation time under the condition $t\rom{exp}>5$ min. The blue filled circle is the result for the fiducial survey setup. Line types are as the same in Fig.~\ref{mnu-tobs}.
} 
\label{fom-tobs}
\end{figure}

Next, in Fig.~\ref{fom-tobs}, similarly to Fig.~\ref{mnu-tobs}, the FoM for various choices of $N$ is plotted against the total observation time $T\rom{obs}$. Note here that varying $t\rom{exp}$, we looked for the maximum value of FoM for each observation time $T\rom{obs}$, under the condition $t\rom{exp}\geq5$ min. Then, the best constraint is obtained from the $N=2$ case, and the improvement of the FoM is roughly proportional to $\sim T\rom{obs}$. However, the resultant constraint is not so drastically changed compared to the one obtained from the canonical setup in Sec.~\ref{sec4}. Even at $T\rom{obs}=250$ nights, the improvement is still below $\sim1.7$. 

The results indicate that joint constraints on the neutrino mass and dark energy equation-of-state parameter is rather stable against the details of the survey setup of the cosmic shear experiment, and in this sense, we could say that the forecast results in Sec.~\ref{sec4} is robust. 

\section{Summary}\label{sec6}

In this paper, we explored a capability and a synergy of the two weak lensing experiments, i.e., CMB lensing and cosmic shear, to simultaneously constrain the neutrino mass and dark energy properties. As representative lensing experiments, we consider the Planck and ACTPol for CMB lensing observations, and Subaru HSC survey for cosmic shear experiment. Including the primary CMB information as a prior cosmological information, the Fisher analysis showed that combining Planck, ACTPol, and HSC, the total mass of neutrinos with $\sim0.1$ eV would be detected with a significance of $\sim 2\sigma$ level. As for the dark energy equation-of-state parameters, the combination of the Planck and HSC would provide a constraint on $w_0$ and $w_a$ with the accuracy of $10\%$ and $20\%$ level, which roughly corresponds to the expected errors for the Stage-III class experiment defined by the Dark Energy Task Force. In other words, including the neutrino mass in the parameter estimation would not drastically alter the FoM estimates of dark energy parameters in weak lensing measurements. The constraints will be further improved by a factor of $\sim2$ if we add the ACTPol data.

We have also studied the role of the cross correlation statistics obtained from the two lensing experiments. While the deflection-shear cross correlation gives a strong statistical correlation between CMB lensing and cosmic shear signals, which makes the improvement of the combined constraints rather moderate, the temperature-deflection cross correlation is found to be sensitive to the early-time evolution of dark energy, and it plays an important role to constrain $w_a$ rather than $w_0$. 

Further, we have investigated the sensitivity to the choice of survey setup specifically focusing on the cosmic shear survey, and compare the results in the canonical setup (Sec.~\ref{sec4}) with those for the optimal survey setup characterized by the exposure time and/or total observation time. Maximizing the efficiency of the survey setup while keeping the total observation time, the optimal number of redshift bins needed for the lensing tomography becomes $N=1$ for neutrino mass and $N=2$ for dark energy equation-of-state parameters, which are contrasted with the canonical setup of $N=3$. However, the resultant improvement for the optimal survey setup is rather small, and the forecast results depend very weakly on the total observation time. Hence, for the setup with a limited observation time, forecast results for the canonical setup presented in Sec.~\ref{sec4} seems robust against the details of the survey setup. 

Note, however, that this conclusion crucially depends on the assumption that the reconstruction of the lensing deflection angle and cosmic shear measurement can be made perfectly without any serious systematics. Further, we assume that the theoretical template for lensing power spectra is well-understood, and can be described with the fitting formula \cite{Smith03}. Also, the non-Gaussian covariance due to the nonlinear gravity has been neglected. These assumptions or treatment would be rather optimistic, and toward proper comparison with observations, the accurate template for lensing spectra should be developed, and the effect of non-Gaussian covariance must be properly incorporated into the parameter estimation analysis \cite{Huterer:2004tr,Takada:2008fn}. In developing these issues, recently proposed analytical techniques to accurately predict matter power spectrum would be helpful (e.g., \cite{Crocce:2007dt,Matsubara:2007wj,Hiramatsu:2009ki,Taruya:2009ir,Pietroni:2008jx,Lawrence:2009uk}), and a large set of N-body simulations is really demanding to study non-Gaussian covariance (e.g., \cite{Sato:2009ct}).  

Throughout the paper, we have focused on the homogeneous dark energy component which only affects the background dynamics of cosmic expansion. However, there may possibly exist a clustering component of dark energy (e.g.\cite{Hu01,HS04}), which can alter the growth of structure on very large scales. As discussed by several authors (e.g.,\cite{Hu01}), CMB lensing and cosmic shear experiments are shown to be complementary probes for dark energy clustering, and even including the effect of free-streaming of massive neutrinos, which appears on relatively small scales, they could still provide fruitful cosmological constraints on both neutrino and dark energy. In this respect, the scientific impacts from the two lensing experiments may become even more large.

Finally, we note that combining the lensing data with other cosmological probes such as the baryon acoustic oscillations and type Ia supernovae can potentially break the degeneracies between $w_0$ and $w_a$ \cite{Tang:2008hm}. Further, the clustering statistics of the galaxy distribution exhibit a free-streaming suppression by the massive neutrinos, and the joint analysis with lensing data would also break the degeneracy between $\mnu$ and $\Om h^2$ \cite{Saito:2009ah,Saito:2010pw}. Although the control of the observational systematics such as galaxy bias and calibration of supernova light curves become very much severe, the scientific impact on the neutrino mass and dark energy parameters would be significant, and a synergy between lensing and other cosmological probes should deserve further investigation. 

\acknowledgments
We thank David Spergel for introducing ACTPol project and Kiyotomo Ichiki, Masahiro Takada, Sudeep Das for helpful discussion. SS and AT are supported in part by a Grants-in-Aid for Scientific Research from the Japan Society for the Promotion of Science (JSPS) (No. 21-00784 for SS and No. 21740168 for AT). Also, SS is supported by JSPS through research fellowships and Excellent Young Researchers Overseas Visit Program. This work was supported in part by Grant-in-Aid for Scientific Research on Priority Areas No. 467 "Probing the Dark Energy through an Extremely Wide and Deep Survey with Subaru Telescope", and JSPS Core-to-Core Program "International Research Network for Dark Energy".

\appendix

\section{Constraints on time varying dark energy equation-of-state parameter} \label{appA}

Sec.~\ref{sec4} reveals that the temperature-deflection cross correlation seems to play an important role to constrain $w_a$, and it helps to break a degeneracy between $\OL$ and $w_a$. In this appendix, employing the principal component analysis (PCA) introduced by Ref.~\cite{HS03}, we investigate the sensitivity of the cross correlation statistics to the time evolution of dark energy in more general context, characterized by the time-varying equation-of-state parameter, $w(z)$. Also, we explore the degeneracy between $\OL$ and $w(z)$, and show that inclusion of the cross correlation helps to break the degeneracy even at high redshifts.

To apply the PCA, let us we first  discretize the equation-of-state parameter as 
\beq
	w(z) = \sum_{i=1}^N w_i\,\Xi(z_{i-1},z_i;z) \,, \label{eq:w_z}
\eeq
where the function $\Xi(z_{i-1},z_i;z)$ is the step function defined by $1$ at $z_{i-1}<z<z_i$, and $0$ otherwise. Here, we set $N=18$ and choose the redshift interval as $z_{i}-z_{i-1}=0.5$. Treating the coefficients $w_i$ as free parameters, we compute the statistical uncertainty of each coefficient from the Fisher matrix in the same way as described in Sec.~\ref{sec3}. Assuming the fiducial values of the coefficients $w_i$ as $-1$ for $i=1,\cdots,N$, we consider the Fisher matrix for the lensing experiments combining Planck, ACTPol and HSC. Then, we obtain the $N\times N$ sub-matrix $\widetilde{\bm{F}}$ for the coefficients $w_i$, marginalized over the other cosmological parameters \footnote{To be precise, we first compute the inverse Fisher matrix, and construct the sub-matrix from it by extracting the components associated with the coefficients $w_i$. Inverting this sub-matrix again, we finally get $\widetilde{\bm{F}}$.}. 

The resultant matrix $\widetilde{\bm{F}}$ generically includes non-vanishing off-diagonal components, which represent statistical correlations between different $w_i$. That is, depending on the survey setup of the lensing experiments, there exist a strong degeneracy between different coefficients, and thereby the constraint on each coefficient $w_i$ cannot be obtained independently. In this sense, a naive discretization of equation-of-state parameter \eqref{eq:w_z} may not be a good description for time-varying equation-of-state parameter. Rather, a better characterization may be obtained by diagonalizing the matrix $\widetilde{\bm{F}}$:  
\beq 
    \widetilde{\bm{F}} = \bm{U}^T\,\bm{\Lambda}\,\bm{U} \,. 
\eeq
The quantity $\bm{\Lambda}$ is the diagonal matrix consisting of the eigenvalues of the matrix $\widetilde{\bm{F}}$, which we denote by $\lambda_j$. The matrix $\bm{U}$ is the orthogonal matrix, whose element of $i$-th row and $j$-th column is given by the eigenvector $\bm{e}_j(z_i)$. Since the eigenvectors $\{\bm{e}_j(z)\}$ can form an orthonormal basis and can be regarded as function of $z$, we may expand the time-varying equation-of-state parameter as
\beq 
    w(z) = \sum_{i=1}^N \alpha_i\,\bm{e}_i(z)
    \,, \label{eq:w_eigen}
\eeq
with the coefficients $\alpha_i$ computed as $\alpha_i=\sum_a w(z_a)\bm{e}_i(z_a)$. In this expansion, the coefficient $\alpha_i$ is statistically independent, and the associated error is inversely proportional to the eigenvalues $\lambda_i$. In this sense, the functions $\bm{e}_i(z)$ are the principal components (PCs), and the redshift dependence of the best-determined eigenvectors characterizes the sensitivity of the measurements to the time evolution of dark energy.  

\begin{figure}
\bc 
\includegraphics[width=13cm,clip]{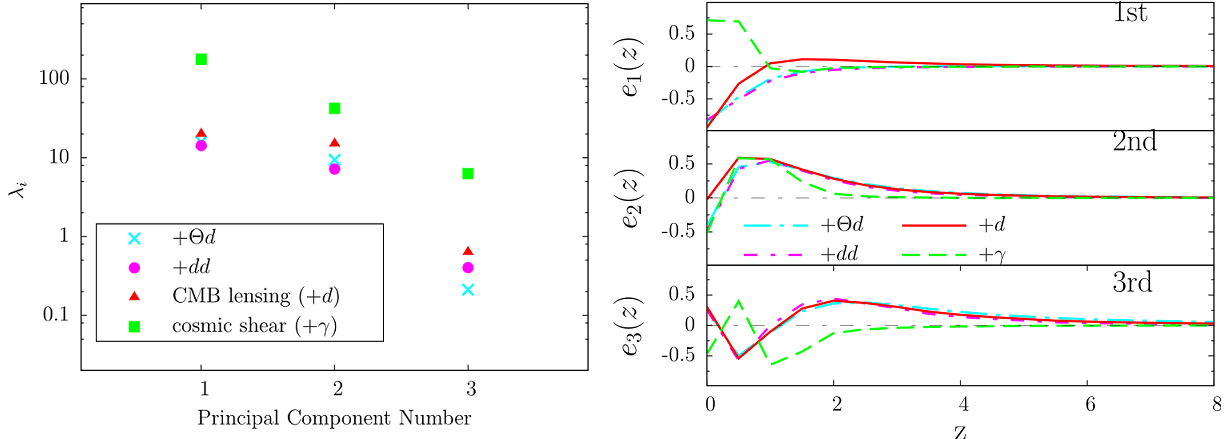}  \vs{-1}
\ec 
\caption{
{\it Left} : The three best-determined eigenvalues $\lambda_i$ for the temperature-deflection cross correlation ($+\Theta d$, cyan cross), deflection auto correlation ($+dd$, magenta filled circle), combining these two signals ($+d$, red filled triangle) and cosmic shear ($+\gamma$, green filled square). {\it Right} : The three best-determined eigenvectors $e_i(z)$ are shown for the temperature-deflection cross correlation ($+\Theta d$: cyan dot-long-dashed), deflection auto correlation ($+dd$: magenta dot-short-dashed), combining these two signals ($+d$: red solid) and cosmic shear ($+\gamma$: green dashed). }
\label{PCA_phi}
\end{figure}

Fig.~\ref{PCA_phi} shows three of the best-determined PCs for eigenvalues (left) and eigenvectors (right). Here, we separately plot the results in four cases, $+\Theta d$, $+dd$, $+d$, and $+\gamma$ (see Secs.~\ref{sec4.2} and \ref{sec4.3}), among which the cosmic shear signal ($+\gamma$) is shown to give the tightest constraint on the dark energy parameters. However, the redshift sensitivity of the cosmic shear is rather restricted to the lower redshift, and the eigenvectors become almost vanishing at $z\gtrsim2$. On the other hand, the eigenvectors associated with the CMB lensing signals (i.e., $+d$,\, $+dd$,\, $+\Theta d$) can have non-vanishing values even at higher redshifts. Although their eigenvalues are basically smaller than those of the cosmic shear, they still have some contributions to the constraints on the dark energy equation-of-state parameters, especially at higher redshifts. An interesting point is that among several CMB lensing signals, temperature-deflection cross correlation seems to give an important contribution to the constraint on the early-time evolution of dark energy, since the first two largest eigenvalues of the temperature-deflection cross correlation are larger than those of the deflection auto correlation. 

\begin{figure}
\bc
\includegraphics[width=9cm]{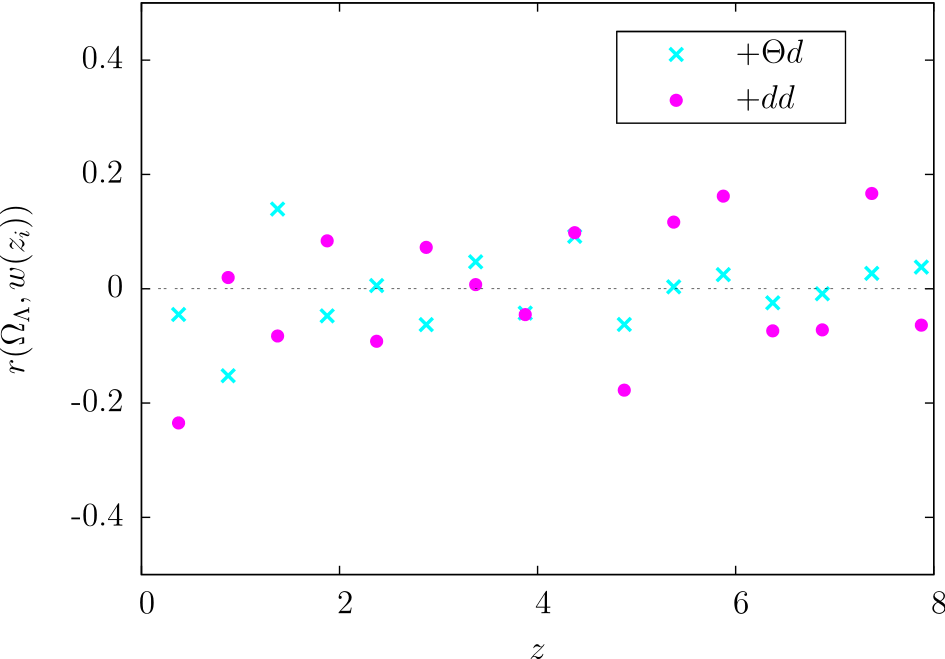}  \vs{-1}
\ec
\caption{
Correlation coefficients $r(\OL,w(z_i))$ for temperature-deflection cross correlation ($+\Theta d$, cyan cross) and the lensing auto correlation ($+dd$, magenta filled circle).
}
\label{PCA_CC}
\end{figure}

To clarify the role of the temperature deflection cross correlation, we calculate the correlation coefficients between $\OL$ and $w_i$, $r(\OL,\,w_i)$, the result of which is shown in Fig.~\ref{PCA_CC}. Note that the parameter $w_i$ implies $w(z_i)$, and the resultant correlation coefficient is plotted against the redshift $z$. As increasing redshift, the absolute value of the correlation coefficients for the temperature-deflection cross correlation becomes smaller than those for the deflection auto correlation. This indicates that the inclusion of temperature-deflection cross correlation helps to break a degeneracy between the density parameter and equation-of-state parameter, especially at higher redshifts. In this respect, the temperature-deflection cross correlation would be a more sensitive probe for the dark energy equation-of-state parameter than the deflection auto correlation.  

\begin{thebibliography}{10}
\def\apj{Astrophys.\ J.\ }
\def\jcap{J.\ Cosmol.\ Astropart.\ Phys.\ }
\def\physrep{Phys.\ Rep.\ }
\def\prd{Phys.\ Rev.\ D }
\def\prl{Phys.\ Rev.\ Lett.\ }
\def\mnras{Mon.\ Not.\ R.\ Astron.\ Soc.\ }

\bibitem{HSC}
Subaru-HSC Collaboration, 
{\it Hyper Suprime-Cam Design Review}, 
2009.

\bibitem{Planck} 
Planck Collaboration, 
{\it Planck: The scientific programme}, 
2006.

\bibitem{Abazajian:2002ck}  
K.~N. Abazajian \& S. Dodelson, 
{\it Neutrino mass and dark energy from weak lensing}, 
\prl \B{91} (2003) 041301.

\bibitem{Alb06}  
A.~Albrecht et~al.,
{\it Report of the Dark Energy Task Force}, 
2006.

\bibitem{Amanullah:2010vv}
R.~Amanullah et~al.,
{\it Spectra and Light Curves of Six Type Ia Supernovae at 0.511 < z < 1.12 and the Union2 Compilation},
\apj \B{716} (2010) 712--738.

\bibitem{AR07}
A. Amara \& A. Refregier,
{\it Optimal Surveys for Weak Lensing Tomography},
\mnras \B{381} (2007) 1018--1026.

\bibitem{BJ02}
G.~M. Bernstein \& M.~Jarvis,
{\it Shapes and Shears, Stars and Smears: Optimal Measurements for Weak Lensing},
Astron. J. \B{123} (2002) 583--618.

\bibitem{BES80}
J.~R. Bond, G.~Efstathiou \& J.~Silk,
{\it Massive neutrinos and the large-scale structure of the universe},
\prl \B{45} (1980) 1980--1984.

\bibitem{Calabrese:2008rt}
E. Calabrese, A. Slosar, A. Melchiorri, G.~F. Smoot \& O. Zahn,
{\it Cosmic Microwave Weak lensing data as a test for the dark universe},
\prd \B{77} (2008) 123531.

\bibitem{Chevallier:2000qy}
M. Chevallier \& D. Polarski,
{\it Accelerating universes with scaling dark matter},
Int. J. Mod. Phys. \B{D10} (2001) 213--224.

\bibitem{Crocce:2007dt}
M. Crocce \& R. Scoccimarro,
{\it Nonlinear Evolution of Baryon Acoustic Oscillations},
\prd \B{77} (2008) 023533.

\bibitem{Das:2010ga}
S. Das et~al.,
{\it The Atacama Cosmology Telescope: A Measurement of the Cosmic Microwave Background Power Spectrum at 148 and 218 GHz from the 2008 Southern Survey},
2010.

\bibitem{DeBernardis:2009di}
F. De~Bernardis, T. D. Kitching, A. Heavens \& A. Melchiorri,
{\it Determining the Neutrino Mass Hierarchy with Cosmology},
\prd \B{80} (2009) 123509.

\bibitem{dP09}
R. de~Putter, O. Zahn \& E.~V. Linder,
{\it CMB Lensing Constraints on Neutrinos and Dark Energy},
\prd \B{79} (2009) 065033.

\bibitem{HTW06}
S. Hannestad, H. Tu \& Y.~Y.~Y. Wong,
{\it Measuring neutrino masses and dark energy with weak lensing tomography},
\jcap \B{06} (2006) 025.

\bibitem{Hiramatsu:2009ki}
T. Hiramatsu \& A. Taruya,
{\it Chasing the non-linear evolution of matter power spectrum with numerical resummation method: solution of closure equations},
\prd \B{79} (2009) 103526.

\bibitem{Hirata08}
C.~M.~Hirata, S. Ho, N. Padmanabhan, U. Seljak \& N.~A. Bahcall,
{\it Correlation of CMB with large-scale structure: II. Weak lensing},
\prd \B{78} (2008) 043520.

\bibitem{HSCS09}
L. Hollenstein, D. Sapone, R. Crittenden \& B.~M. Schaefer,
{\it Constraints on early dark energy from CMB lensing and weak lensing tomography},
\jcap \B{04} (2009) 012.

\bibitem{Hu99}
W. Hu, 
{\it Power Spectrum Tomography with Weak Lensing},
\apj \B{522} (1999) L21--L24.

\bibitem{Hu01}
W. Hu,
{\it Dark Synergy: Gravitational Lensing and the CMB},
\prd \B{65} (2002) 023003.

\bibitem{HO02}
W. Hu \& T. Okamoto,
{\it Mass Reconstruction with CMB Polarization},
\apj \B{574} (2002) 566--574.

\bibitem{HS04}
W. Hu \& R. Scranton,
{\it Measuring Dark Energy Clustering with CMB-Galaxy Correlations},
\prd \B{70} (2004) 123002.

\bibitem{HS03}
D. Huterer \& G. Starkman,
{\it Parameterization of dark-energy properties: A principal- component approach},
\prl \B{90} (2003) 031301.

\bibitem{Huterer:2004tr}
D. Huterer \& M. Takada,
{\it Calibrating the Nonlinear Matter Power Spectrum: Requirements for Future Weak Lensing Surveys},
Astropart. Phys. \B{23} (2005) 369--376.

\bibitem{I09}
K. Ichiki, M. Takada \& Tomo Takahashi,
{\it Constraints on Neutrino Masses from Weak Lensing},
\prd \B{79} (2009) 023520.

\bibitem{Kaplinghat:2003bh}
M. Kaplinghat, L. Knox \& Y.-S. Song,
{\it Determining neutrino mass from the CMB alone},
\prl \B{91} (2003) 241301.

\bibitem{Kitching:2008dp}
T.~D. Kitching, A.~F. Heavens, L.~Verde, P.~Serra, and A.~Melchiorri,
{\it Finding Evidence for Massive Neutrinos using 3D Weak Lensing},
\prd \B{77} (2008) 103008.

\bibitem{Komatsu:2010fb}
E.~Komatsu et~al.,
{\it Seven-Year Wilkinson Microwave Anisotropy Probe (WMAP) Observations: Cosmological Interpretation},
2010.

\bibitem{Lawrence:2009uk}
E. Lawrence et~al.,
{\it The Coyote Universe III: Simulation Suite and Precision Emulator for the Nonlinear Matter Power Spectrum},
\apj \B{713} (2010) 1322--1331.

\bibitem{Lesgourgues:2006nd}
J. Lesgourgues \& S. Pastor,
{\it Massive neutrinos and cosmology},
\physrep \B{429} (2006) 307--379.

\bibitem{LPPP06}
J. Lesgourgues, L. Perotto, S. Pastor \& Michel Piat,
{\it Probing neutrino masses with CMB lensing extraction},
\prd \B{73} (2006) 045021.

\bibitem{LC06}
A. Lewis \& A. Challinor,
{\it Weak Gravitational Lensing of the CMB},
\physrep {429} (2006) 1--65.

\bibitem{CAMB}
A. Lewis, A. Challinor \& A. Lasenby,
{\it Efficient Computation of CMB anisotropies in closed FRW models},
\apj \B{538} (2000) 473--476.

\bibitem{MSTV04}
M.~Maltoni, T.~Schwetz, M.~A. Tortola \& J.~W.~F. Valle, 
{\it Status of global fits to neutrino oscillations},
New J. Phys. \B{6} (2004) 122.

\bibitem{Matsubara:2007wj}
T. Matsubara,
{\it Resumming Cosmological Perturbations via the Lagrangian Picture: One-loop Results in Real Space and in Redshift Space},
\prd \B{77} (2008) 063530.

\bibitem{Munshi:2006fn}
D.~Munshi, P.~Valageas, Ludovic Van~Waerbeke \& A.~Heavens,
{\it Cosmology with Weak Lensing Surveys},
\physrep \B{462} (2008) 67--121.

\bibitem{Niemack:2010wz}
M.~D. Niemack et~al.,
{\it ACTPol: A polarization-sensitive receiver for the Atacama Cosmology Telescope},
Proc. SPIE Int. Soc. Opt. Eng. \B{7741} (2010) 77411S.

\bibitem{OH03}
T. Okamoto \& W. Hu,
{\it CMB Lensing Reconstruction on the Full Sky},
\prd \B{67} (2003) 083002.

\bibitem{Perl99}
S.~Perlmutter et~al.,
{\it Measurements of Omega and Lambda from 42 High-Redshift Supernovae},
\apj \B{517} (1999) 565--586.

\bibitem{Perotto:2006rj}
L. Perotto, J. Lesgourgues, S. Hannestad, H. Tu \& Y.~Y.~Y. Wong,
{\it Probing cosmological parameters with the CMB: Forecasts from full Monte Carlo simulations},
\jcap \B{10} (2006) 013.

\bibitem{Pietroni:2008jx}
M. Pietroni,
{\it Flowing with Time: a New Approach to Nonlinear Cosmological Perturbations},
\jcap \B{10} (2008) 036.

\bibitem{Reid:2009nq}
B.~A. Reid, L. Verde, R. Jimenez \& O. Mena,
{\it Robust Neutrino Constraints by Combining Low Redshift Observations with the CMB},
\jcap \B{01} (2010) 003.

\bibitem{Riess98}
A.~G. Riess et~al.,
{\it Observational Evidence from Supernovae for an Accelerating Universe and a Cosmological Constant},
Astron. J. \B{116} (1998) 1009--1038.

\bibitem{Saito:2009ah}
S. Saito, M. Takada \& A. Taruya,
{\it Nonlinear power spectrum in the presence of massive neutrinos: perturbation theory approach, galaxy bias and parameter forecasts},
\prd \B{80} (2009) 083528.

\bibitem{Saito:2010pw}
S. Saito, M. Takada \& A. Taruya,
{\it Neutrino mass constraint with SDSS LRG power spectrum and perturbation theory},
2010.

\bibitem{Sato:2009ct}
M. Sato et~al.,
{\it Simulations of Wide-Field Weak Lensing Surveys I: Basic Statistics and Non-Gaussian Effects},
\apj \B{701} (2009) 945--954.

\bibitem{Smith:2007rg}
K.~M. Smith, O. Zahn \& Olivier Dore,
{\it Detection of Gravitational Lensing in the Cosmic Microwave Background},
\prd \B{76} (2007) 043510.

\bibitem{Smith03}
R.~E. Smith et~al.,
{\it Stable clustering, the halo model and nonlinear cosmological power spectra},
\mnras \B{341} (2003) 1311.

\bibitem{SK04}
Y.-S. Song \& L. Knox,
{\it The detectability of departures from the inflationary consistency equation},
\prd \B{68} (2003) 043518.

\bibitem{ACTPol}
D.~N. Spergel, (private comunication).

\bibitem{Takada:2008fn}
M. Takada \& B. Jain,
{\it The Impact of Non-Gaussian Errors on Weak Lensing Surveys},
\mnras \B{395} (2009) 2065.

\bibitem{Takahashi:2009ty}
R. Takahashi et~al.,
{\it Non-Gaussian Error Contribution to Likelihood Analysis of the Matter Power Spectrum},
\apj \B{726} (2011) 7.

\bibitem{Tang:2008hm}
J. Tang, F.~B. Abdalla \& J. Weller,
{\it Complementarity of Future Dark Energy Probes},
2008.

\bibitem{Taruya:2009ir}
A. Taruya, T. Nishimichi, S. Saito \& T. Hiramatsu,
{\it Non-linear Evolution of Baryon Acoustic Oscillations from Improved Perturbation Theory in Real and Redshift Spaces},
\prd \B{80} (2009) 123503.

\bibitem{TTH97}
M. Tegmark, A. Taylor \& A. Heavens,
{\it Karhunen-Loeve eigenvalue problems in cosmology: how should we tackle large data sets ?},
\apj \B{480} (1997) 22.

\bibitem{Verde:2005ff}
L. Verde, H. Peiris \& R. Jimenez,
{\it Optimizing CMB polarization experiments to constrain inflationary physics},
\jcap \B{01} (2006) 019.

\bibitem{YPHNS07}
K. Yamamoto, D. Parkinson, T. Hamana, R.~C. Nichol \& Y. Suto,
{\it Optimizing future imaging survey of galaxies to confront dark energy and modified gravity models},
\prd \B{76} (2007) 023504.

\bibitem{Zaldarriaga:1998ar}
M. Zaldarriaga \& U. Seljak,
{\it Gravitational Lensing Effect on Cosmic Microwave Background Polarization},
\prd \B{58} (1998) 023003.

\end{thebibliography}
\end{document}